\documentclass[journal,12pt,onecolumn,draftclsnofoot,]{IEEEtran}
\IEEEoverridecommandlockouts
\ifCLASSINFOpdf
  \usepackage[pdftex]{graphicx}
\else
\fi
\usepackage[cmex10]{amsmath}
\usepackage{array}
\usepackage{amsthm}
\usepackage{amsmath}
\usepackage{amsfonts}
\usepackage{balance}

\newtheorem{definition}{Definition}
\newtheorem{theorem}{Theorem}

\newtheorem{lemma}{Lemma}
\newtheorem{proposition}{Proposition}

\allowdisplaybreaks

\begin{document}
\title{Source Coding in Networks with Covariance Distortion Constraints}

\author{Adel~Zahedi,
        Jan~\O stergaard,~\IEEEmembership{Senior Member,~IEEE,}
        S\o ren~Holdt~Jensen,~\IEEEmembership{Senior Member,~IEEE,}
        Patrick~A.~Naylor,~\IEEEmembership{Senior Member,~IEEE,}        
        and~S\o ren~Bech%,~\IEEEmembership{Life~Fellow,~IEEE}% <-this % stops a space
\thanks{The research leading to these results has received funding from the European Union's Seventh Framework Programme (FP7/2007-2013) under grant agreement n$^\circ$ ITN-GA-2012-316969. The work of J. \O stergaard is financially supported by VILLUM FONDEN Young Investigator Programme, Project No. 10095. This work was presented in part at the 2014 IEEE International Symposium on Information Theory \cite{isit} and the IEEE Data Compression Conference \cite{DCC15}.

Copyright (c) 2016 IEEE. Personal use of this material is permitted. However, permission to use this material for any other purposes must be obtained from the IEEE by sending a request to pubs-permissions@ieee.org.

A. Zahedi, J. \O stergaard, and S. H. Jensen (email: \{adz, jo, shj\}@es.aau.dk) are with the Department of Electronic Systems, Aalborg University, Denmark. P. Naylor (email: p.naylor@imperial.ac.uk) is with the Electrical and Electronic Engineering Department, Imperial College London, UK, and S. Bech (email: sbe@es.aau.dk) is with Bang \& Olufsen, Denmark and the Department of Electronic Systems, Aalborg University, Denmark.}% <-this % stops a space
%\thanks{Manuscript received April 19, 2005; revised December 27, 2012.}
}

\maketitle

\begin{abstract}
We consider a source coding problem with a network scenario in mind, and formulate it as a remote vector Gaussian Wyner-Ziv problem under covariance matrix distortions. We define a notion of minimum for two positive-definite matrices based on which we derive an explicit formula for the rate-distortion function (RDF). We then study the special cases and applications of this result. We show that two well-studied source coding problems, i.e. remote vector Gaussian Wyner-Ziv problems with mean-squared error and mutual information constraints are in fact special cases of our results. Finally, we apply our results to a joint source coding and denoising problem. We consider a network with a centralized topology and a given weighted sum-rate constraint, where the received signals at the center are to be fused to maximize the output SNR while enforcing no linear distortion. We show that one can design the distortion matrices at the nodes in order to maximize the output SNR at the fusion center. We thereby bridge between denoising and source coding within this setup.

\end{abstract}


% \fi
\IEEEpeerreviewmaketitle

%********************************************************************************************************************************************************
%********************************************************************************************************************************************************
%********************************************************************************************************************************************************
%********************************************************************************************************************************************************

\section{Introduction}
\label{intro}

\begin{figure}[!t]
\centering
\includegraphics[width=3in]{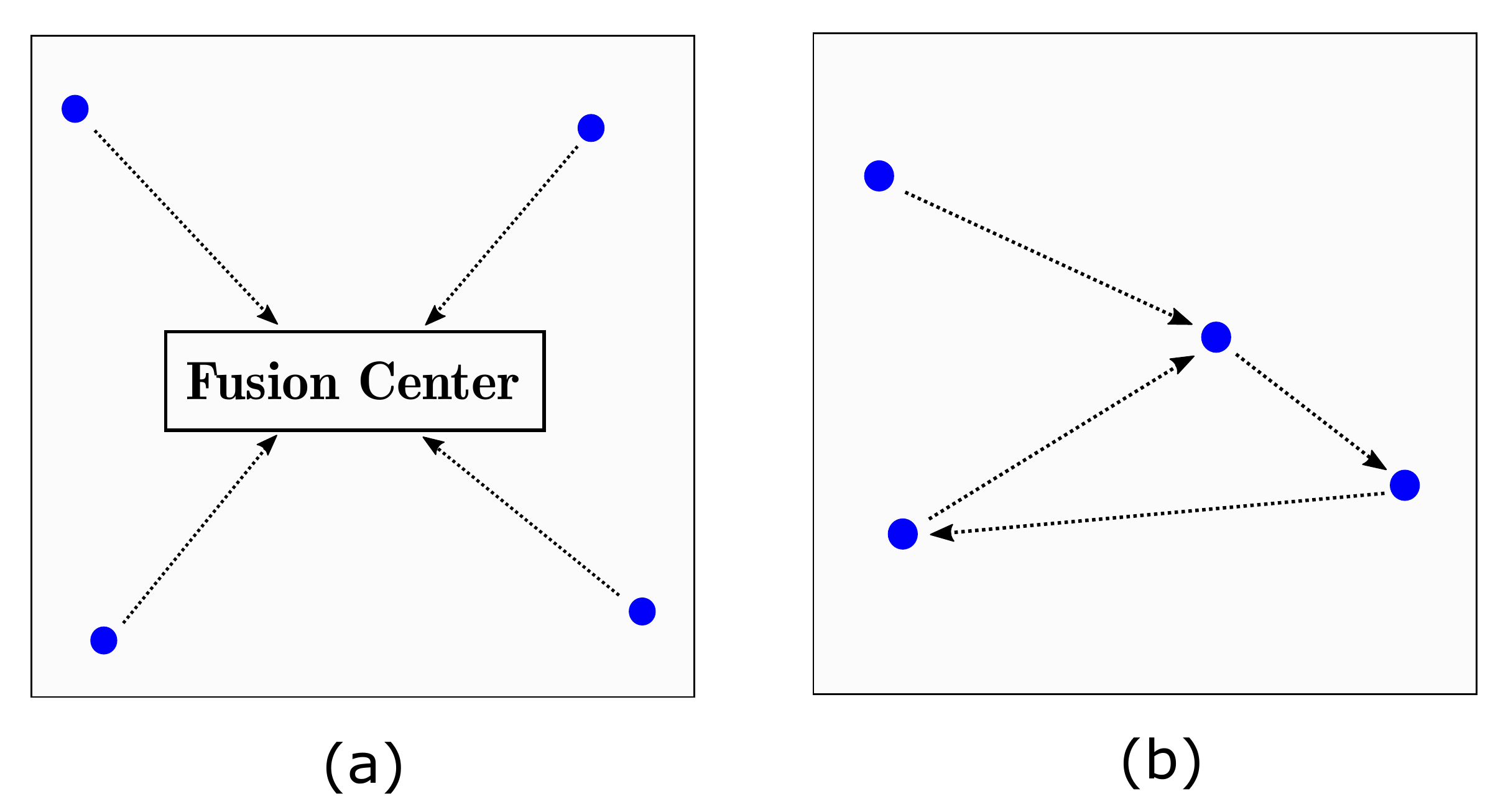}
\caption{Examples of wireless sensor networks (a) with a centralized topology, (b) with an ad hoc topology}
\label{topology}
\end{figure}

\begin{figure}[!t]
\centering
\includegraphics[width=3.5in]{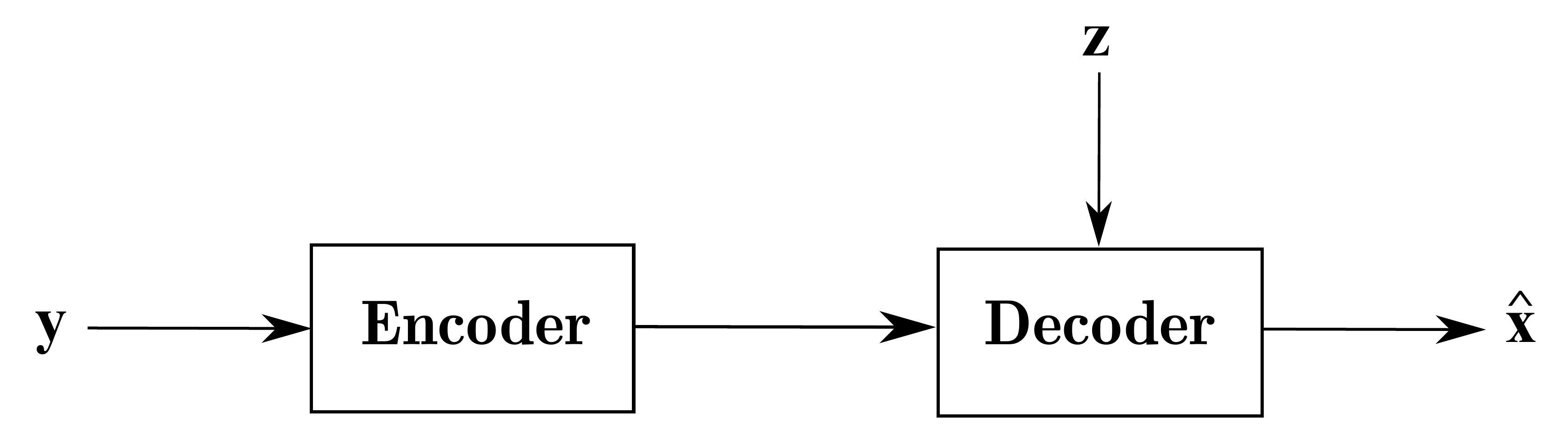}
\vspace{-2 mm}
\caption{Block diagram of the remote source coding problem}
\label{fig}
\vspace{-2 mm}
\end{figure}

It is an inherent property of sensor networks that there are several observations of the desired source due to the availability of several sensors. If the observations at different nodes are correlated, it makes sense to make use of the correlation among the measurements to reduce the communication costs. This is particularly important for wireless networks where the resources, i.e. power and bandwidth are limited. In networks with a centralized topology, as shown in Fig. \!\ref{topology} \!(a), the measurements from all the nodes are to be sent directly to a fusion center. In this case, one could think of the data that is already available at the fusion center (sent by other nodes) as side information. For an ad hoc topology with communication among the nodes, as shown in Fig. \!\!\ref{topology} \!(b), the observation at the receiving node could be treated as side information. In either case, due to the correlation between the side information at the receiver and the data to be sent, the required transmission rate could be reduced using distributed source coding \cite{DSC,SP}. The block diagram of the resulting source coding problem with side information is shown in Fig. \!\ref{fig}. The vectors ${\bf{x}}$, ${\bf{y}}$, and ${\bf{z}}$ are the desired source, the observation at the encoder, and the side information available to the decoder, respectively. As shown in Fig. \!\ref{fig}, an optimal estimation $\hat{\bf{x}}$ of ${\bf{x}}$ is obtained from the received data and the side information at the decoder. This setup is commonly referred to as the Wyner-Ziv problem after the celebrated work by Wyner and Ziv \cite{WZ}. Note that this setup is less general than those shown in Fig. \!1, where there may be several sources. However, we focus on this setup for simplicity.

In this work, we consider i.i.d. vector sources. Note that in practice, the sources are correlated in time. By considering i.i.d. vector sources, we allow for correlation within each vector. This partially models the memory in the sources. To make full use of the memory in the signals, one has to either avoid the i.i.d. assumption, or let the dimension of the vector sources tend to infinity. For our derivations, we use the i.i.d. assumption and finite dimensions, and therefore accept some level of suboptimality.

The Wyner-Ziv problem first appeared as an attempt to generalize the Slepian and Wolf's results in \cite{SW} for lossless coding of two digital sources to the case of lossy source coding. The main result in \cite{SW} was that, it is possible to encode two digital sources losslessly and separately with a sum-rate that asymptotically achieves the joint entropy of the two sources given that the decoding is performed jointly. In \cite{WZ}, Wyner and Ziv considered a similar problem for lossy source coding, where one source is to be encoded and sent to the decoder and the other one is available only to the decoder and serves as the side information. The results were later extended to continuous sources in \cite{Wyner2}. It was shown that in general, this setup incurs a loss compared to the case where the side information is available at the encoder, but if the sources are Gaussian, no loss occurs. 

Since in our setup the observation ${\bf{y}}$ is in general not the same as the desired source ${\bf{x}}$, we refer to it as the remote Wyner-Ziv problem. In this paper, we consider the remote Wyner-Ziv problem with a covariance matrix distortion constraint. We derive a closed-form formula for the rate-distortion function (RDF), and then study the special cases and applications of this result to source coding problems in networks. In order to formally define the problem, we introduce the distortion constraints which are of interest for this work in Section \ref{define-distortions}, but before that, we need to define our notation.

%********************************************************************************************************************************************************
\subsection{Notation}
\label{not}

We denote matrices and vectors by boldface uppercase and lowercase letters, respectively. We consider zero-mean stationary Gaussian sources, which generate independent vectors ${\bf{x}} \in {\mathbb{R}}^{n_x}$, ${\bf{y}} \in {\mathbb{R}}^{n_y}$, and ${\bf{z}} \in {\mathbb{R}}^{n_z}$, as shown in Fig. \!\ref{fig}. The ordered set of vectors $({\bf{x}}_1,{\bf{x}}_2,\cdots,{\bf{x}}_N)$ is denoted by $\{ {\bf{x}}_i\}_{i=1}^N$ for compactness. We denote the expectation operation by $E[\cdot]$. The covariance and cross-covariance matrices are denoted by the symbol ${\bf{\Sigma}}$ followed by an appropriate subscript. As an example, for random vectors ${\bf{x}}$, ${\bf{y}}$ and ${\bf{z}}$, the conditional cross-covariance of ${\bf{x}}$ and ${\bf{y}}$ given ${\bf{z}}$ and the conditional covariance of ${\bf{x}}$ given ${\bf{y}}$ and ${\bf{z}}$ are denoted by ${{\bf{\Sigma }}_{{\bf{x}}{{\bf{y}}}|{{\bf{z}}}}}$ and ${{\bf{\Sigma }}_{{\bf{x}}|{{\bf{y}}}{{\bf{z}}}}}$, respectively, and are defined as:

\begin{align*}
{{\bf{\Sigma }}_{{\bf{x}}{{\bf{y}}}|{{\bf{z}}}}} & \buildrel \Delta \over = E \left[ ({\bf{x}} - {\boldsymbol{\mu}}_{{\bf{x}}|{\bf{z}}})({\bf{y}} - {\boldsymbol{\mu}}_{{\bf{y}}|{\bf{z}}})^T \left| {\bf{z}} \right], \right. \\
{{\bf{\Sigma }}_{{\bf{x}}|{{\bf{y}}}{{\bf{z}}}}} & \buildrel \Delta \over = E \left[ ({\bf{x}} - {\boldsymbol{\mu}}_{{\bf{x}}|{\bf{y}},{\bf{z}}})({\bf{x}} - {\boldsymbol{\mu}}_{{\bf{x}}|{\bf{y}},{\bf{z}}})^T \left| {\bf{y}},{\bf{z}} \right], \right.
\end{align*}

\noindent where the superscript $^T$ denotes the transpose of a matrix, and $\boldsymbol{\mu}$ denotes the mean. For instance, ${\boldsymbol{\mu}}_{{\bf{x}}|{\bf{y}},{\bf{z}}}= E[{\bf{x}}|{\bf{y}},{\bf{z}}]$ is the conditional mean of ${\bf{x}}$ given ${\bf{y}}$ and ${\bf{z}}$. 

Markov chains are denoted by two-headed arrows as in ${\bf{x}} \leftrightarrow {\bf{y}} \leftrightarrow {\bf{u}}$, which means that if ${\bf{y}}$ is given, then ${\bf{x}}$ and ${\bf{u}}$ are independent. The information theoretic operations of differential entropy and mutual information are denoted by $h(\cdot)$ and $I(\cdot;\cdot)$, respectively. The trace operation is denoted by $\text{tr}(\cdot)$. The set of symmetric positive-definite matrices is denoted by ${\mathbb{S}}^+$. The statement ${\bf{A} \succeq {\bf{B}}}$ (${\bf{A} \succ {\bf{B}}}$) means that ${\bf{A} - {\bf{B}}}$ is positive semidefinite (definite). The $n \times n$ identity matrix is denoted by ${\bf{I}}_n$. We use $\text{diag}\{\lambda_i, i=1,\dotsc,n\}$ to show an $n \times n$ diagonal matrix having the elements $\lambda_1, \dotsc, \lambda_n$ on its main diagonal. We also make use of the following notations for briefness:

\begin{align*}
{(a)^ + } \buildrel \Delta \over = \max \left( {a,1} \right), \,\,\,\,\,\, {(a)^ - } \buildrel \Delta \over = \min \left( {a,1} \right).
\end{align*}

%********************************************************************************************************************************************************
\subsection{Distortion Constraints}
\label{define-distortions}

Consider the remote Wyner-Ziv problem where the decoder makes an estimation of the desired source ${\bf{x}} \in {\mathbb{R}}^{n_x}$ using the received message ${\bf{w}} \in {\mathbb{R}}^{n_w}$ and the side information ${\bf{z}} \in {\mathbb{R}}^{n_z}$. For the special case where the sources are scalar, the well-known mean-squared error distortion constraint is in the form of $\sigma^2_{x|wz} \leq D$, where $D$ is a given target distortion, and $\sigma^2_{x|wz}$ is the variance of the reconstruction error. With this distortion constraint, the scalar Gaussian remote Wyner-Ziv problem was treated in \cite{Wornell}--\cite{ieice}, and the rate-distortion function was derived in a rather simple form. 

For the vector Gaussian case, there are several options to define a quadratic distortion constraint. The most obvious one is the mean-squared error distortion constraint, which is the generalization of the above-mentioned distortion constraint for the case of vector sources. It is defined as follows:

\begin{align*}
\text{tr} \left( {{\bf{\Sigma }}_{{\bf{x}}|{{\bf{w}}}{{\bf{z}}}}} \right) \leq n_x D,
\end{align*}

\noindent where $D$ is a given target distortion. Another constraint, which is related to a relay network problem as discussed in Section \ref{mutual_info}, is a mutual information constraint defined as:

\begin{align*}
I \left( {{\bf{x}};{\bf{w}}|{{\bf{z}}}} \right) \!\geq\! R_I
\end{align*}

\noindent where $R_I$ is a given constant. For both the mean-squared error distortion and mutual information constraints, the vector Gaussian problem was solved in \cite{Tian}, and turned out to have a parametric form, resembling the familiar water-filling solutions for some other Gaussian rate-distortion problems.

The last constraint we discuss is the covariance matrix distortion constraint, which is defined as:

\begin{equation}
\label{covariance_distortion_constraint}
{{\bf{\Sigma }}_{{\bf{x}}|{{\bf{w}}}{{\bf{z}}}}} \preceq {\bf{D}},
\end{equation}

\noindent where ${\bf{D}}$ is a given positive-definite target distortion matrix. For the vector Gaussian Wyner-Ziv problem with the covariance matrix distortion constraint, the rate-distortion function can be derived using standard techniques for a limited set of distortion matrices (e.g. ${\bf{D}} \preceq {{\bf{\Sigma }}_{\bf{x|z}}}$) \cite{isit}. However, the general case does not appear to be manageable to solve using standard techniques, and no closed-form statement is available for the general RDF in the literature. In this paper, we treat this problem and study some of its applications. We elaborate on some aspects of this problem and our motivation in the next subsection.

%********************************************************************************************************************************************************
\subsection{Motivation}
During the recent years, there has been a shift from the traditional mean-squared error distortion constraint to covariance matrix distortions in the area of multiterminal source coding \cite{Hua-Wang}--\cite{Oohama}. 

In \cite{Hua-Wang}, a multiple description scenario is considered where $L$ encoders send their descriptions of a vector Gaussian source to $L$ individual and one joint decoders. Each decoder makes an estimation of the source within a given covariance matrix distortion constraint. The sum-rate of the multiple descriptions problem is specified in \cite{Hua-Wang} under certain assumptions on the distortion matrices. 

The Gaussian one-helper problem with the covariance matrix distortion constraint was studied in \cite{Wagner2012,Wagner2015}. Here in addition to a main encoder which observes the direct source and sends a message to the decoder with rate $R_1$, there is a helper with noisy observations, which sends a message with rate $R_2$ to the decoder. The decoder estimates the source using the received messages. The rate regions for this problem for the scalar and general helpers were derived in \cite{Wagner2012} and \cite{Wagner2015}, respectively.

In addition to the above works, there are also several results which bound the rate-distortion regions of source coding problems under the covariance matrix distortion constraint \cite{multi_terminal}--\cite{Oohama}. In the so-called CEO problem \cite{CEO}, a number of \emph{agents} encode and send their observations of a source with certain rates to a \emph{CEO}. The CEO makes an estimation of the source using the received messages from the agents. The rate region for the vector Gaussian CEO problem under the covariance matrix distortion constraint was outer-bounded in \cite{multi_terminal,Ekrem}. Finally, in \cite{Oohama}, a set of $L$ noisy observations of $K$ correlated scalar Gaussian sources are encoded separately and sent with the rates $R_1,\cdots,R_L$ to a joint decoder. The decoder makes an estimation of the sources using the received messages. The rate-distortion regions for this problem are inner- and outer-bounded for three different distortion constraints including the covariance matrix distortions.

Although the vector Gaussian remote Wyner-Ziv problem has a rather simple setup, we could not find any closed-form solution for the general problem under the covariance matrix distortion constraint in the literature. Neither does such a closed-form solution follow as a special case of the above-mentioned results.

In general, with the covariance matrix distortion constraint, the matrix form of the target distortion gives rise to new issues compared to the scalar target distortions, which make the problem harder to solve. For the Wyner-Ziv problem, the covariance matrix of the unknown part of the source at the decoder is ${{\bf{\Sigma }}_{{\bf{x}}|{{\bf{z}}}}}$. With a scalar target distortion $d$, the following two cases might arise: 

\begin{itemize}
\item $\text{tr}({{\bf{\Sigma }}_{{\bf{x}}|{{\bf{z}}}}}) \leq n_xd$, for which the required rate is zero.
\item $\text{tr}({{\bf{\Sigma }}_{{\bf{x}}|{{\bf{z}}}}}) > n_xd$, which means that some nonzero rate has to be spent. In this case, with an optimal coding scheme, the variance of the reconstruction error at the decoder would be the same as the target distortion.
\end{itemize}

\noindent
For a matrix target distortion $\bf{D}$ on the other hand, it might happen that none of the two cases ${{\bf{\Sigma }}_{{\bf{x}}|{{\bf{z}}}}} \preceq \bf{D}$ and ${{\bf{\Sigma }}_{{\bf{x}}|{{\bf{z}}}}} \succeq \bf{D}$ hold. In general, there are two sets of directions in ${\mathbb{R}}^{n_x}$. In one set of directions, the distortion constraint is already satisfied at the decoder using the side information ${{\bf{z}}}$. In the other set of directions, the distortion constraint cannot be satisfied without some help from the encoder, making the minimum required rate a nonzero value. In such cases, the covariance matrix of the reconstruction error at the decoder is not guaranteed to be equal to $\bf{D}$.

The argument for considering the covariance matrix distortion constraints despite the above-mentioned issues is the generality of the results. For the remote Wyner-Ziv problem, the resulting rate-distortion function $R(\bf{D})$ would be a mapping from ${\mathcal{D}} = \left\{ {\bf{D}} \in {\mathbb{S}}^+ | \, {\bf{D}} \succ {{\bf{\Sigma }}_{{\bf{x}}|{{\bf{yz}}}}} \right\}$ to ${\mathbb{R}}$. For a given problem with a constraint on ${\bf{D}}$, there is in general a subset of ${\mathcal{D}}$ for which the constraint is satisfied. We can then choose one member of this subset which minimizes the rate. As an example (among other examples that will be studied in this paper), we will show later on, that the ubiquitous case of the RDF with the mean-squared error distortion is simply equivalent to $R(\bf{D})$ at a particular distortion matrix $\bf{D}^*$ in ${\mathcal{D}}$. This particular $\bf{D}^*$ is the one that minimizes the rate $R({\bf{D}})$ subject to a mean-squared error constraint on $\bf{D}$.

In addition to generality, the covariance matrix distortion constraint has other advantages. Defining the distortion with respect to the covariance matrix of the reconstruction error allows for formulating new problems. In many sensor network setups, there are constraints on the sum-rate or weighted sum-rate of the network. As we will show in Section \ref{applications}, this can be formulated as a constraint on the determinant of the distortion matrices. One can then optimize a function (such as SNR) over the network subject to this constraint in order to optimally allocate the rates and distortions to the sensors.

%********************************************************************************************************************************************************
\subsection{Overview of the Paper}
Section \ref{WZ} is dedicated to the vector Gaussian remote Wyner-Ziv problem under covariance matrix distortion constraints. We start with defining a minimum for a pair of symmetric positive-definite matrices based on the joint diagonalization of the matrices. We then derive some properties for this notion of minimum of two matrices, based on which we find an explicit formula for $R(\bf{D})$. This minimum of two matrices seems to be natural to our source coding problem, and in addition to the RDF itself, it appears also in the coding schemes and the reconstruction error at the decoder. To derive the RDF, we first lower-bound $R(\bf{D})$ using the properties of the minimum of two matrices and information-theoretic arguments. Next we upper-bound it by suggesting a linear coding scheme. The lower and upper bounds coincide, thus yielding the RDF.

In Section \ref{applications}, we present examples of applications and special cases of the results obtained in Section \ref{WZ}. We consider three applications. First, we consider a similar source coding problem under the mean-squared error distortion constraint, and will show that the resulting RDF is a special case of $R(\bf{D})$ for a certain choice of the distortion matrix $\bf{D}$. 

Next, we consider a relay network scenario where in addition to the data transmitted to the center from a main transmitter, there is a relay which would like to help the main transmitter by sending its own observation to the center with a certain rate. The problem could be formulated as a source coding problem giving a rate-information function \cite{Tian}. We will show that similar to the previous case, the resulting rate-information function will be given by $R(\bf{D})$ for a certain choice of the distortion matrix $\bf{D}$. We then show how one could implement this scenario using the results from Section \ref{WZ}.

Finally, as the third and most elaborated example of the applications, we consider a sensor network with a centralized topology. The observation from each sensor is encoded and transmitted to the fusion center with a certain rate. We assume that the the weighted sum-rate of the network is limited to a given amount. At the fusion center one would like to fuse the received data in a manner that the output experiences no linear distortion. We will show that under the given weighted sum-rate constraint, one could design and allocate the distortion matrices to the sensor nodes in order to maximize the output signal to noise ratio at the fusion center.

We conclude the paper in Section \ref{conclusion} by discussing the main results and possibilities for future work.

%********************************************************************************************************************************************************
%********************************************************************************************************************************************************
%********************************************************************************************************************************************************
%********************************************************************************************************************************************************

\section{Vector Gaussian Remote Wyner-Ziv Problem}
\label{WZ}

In this section, we solve the source coding problem in networks which was simplified to the block diagram in Fig. \ref{fig}. Later in Section \ref{applications}, we will show how we can use the results for optimizing certain functions in networks. We start with a formal definition of the source coding problem that was introduced in Section \ref{intro}. Assume that $\{ {\bf{x}}_i,{\bf{y}}_i,{\bf{z}}_i \}_{i=1}^N$ is a sequence of i.i.d. zero-mean random vectors such that ${\bf{x}}_i \in {\mathbb{R}}^{n_x}$, ${\bf{y}}_i \in {\mathbb{R}}^{n_y}$, and ${\bf{z}}_i \in {\mathbb{R}}^{n_z}$ are jointly Gaussian for $i=1,\cdots, N$. The encoder observes $\{ {\bf{y}}_i\}_{i=1}^N$, and using an encoding function:

\begin{equation*}
\phi^{(N)}: {\mathbb{R}}^{n_yN} \rightarrow \lbrace 1,\cdots,M^{(N)} \rbrace,
\end{equation*}

\noindent
sends a message to the decoder. The decoder observes $\{ {\bf{z}}_i\}_{i=1}^N$ and receives the message from the encoder, based on which makes an estimation of $\{ {\bf{x}}_i\}_{i=1}^N$ using a decoding function:

\begin{equation*}
\psi^{(N)}: \lbrace 1,\cdots,M^{(N)} \rbrace \times {\mathbb{R}}^{n_zN} \rightarrow {\mathbb{R}}^{n_xN}.
\end{equation*}

\begin{definition}
The rate-distortion pair $(R,{\bf{D}})$ is achievable if there exists a block length $N$ and the encoding and decoding functions $\phi^{(N)}$ and $\psi^{(N)}$, such that for $\{ \hat{\bf{x}}_i\}_{i=1}^N = \psi^{(N)} \left( \phi^{(N)} (\{ {\bf{y}}_i\}_{i=1}^N) , \{ {\bf{z}}_i\}_{i=1}^N \right)$:

\begin{align*}
& R \geq \frac{1}{N} \log{ M^{(N)}}, \\
& {\bf{D}} \succeq \frac{1}{N} \sum_{i=1}^N E\left[ \left( {\bf{x}}_i - \hat{\bf{x}}_i \right) \left( {\bf{x}}_i - \hat{\bf{x}}_i \right)^T \right].
\end{align*}

\end{definition}

\begin{definition}
The rate-distortion region for the above problem is the closure of the set of achievable rate-distortion pairs $(R,{\bf{D}})$.
\end{definition}

\begin{definition}
Let $\mathcal{R}$ be the rate-distortion region for the above problem. The RDF $R({\bf{D}})$ is then defined as:

\begin{equation*}
R({\bf{D}}) = \inf_{(R,{\bf{D}}) \in \mathcal{R}} R.
\end{equation*}
\end{definition}

Notice that ${{\bf{\Sigma }}_{\bf{x|yz}}}$ is the covariance matrix of the reconstruction error at the decoder when in addition to the side information ${\bf{z}}$, the uncoded observation ${\bf{y}}$ of the encoder is also available at the decoder. Since this is not achievable with finite rates, we must have ${{\bf{\Sigma }}_{\bf{x|yz}}} \prec \bf{D}$, otherwise the rate would become infinite. We thus assume that it holds.

Suppose that the auxiliary random variable ${\bf{u}}$ satisfies the Markov chain ${\bf{u}} \leftrightarrow {\bf{y}} \leftrightarrow ({\bf{x}},{\bf{z}})$. From the results in \cite{Wyner2}, it follows that the operational RDF defined above is equal to the following information RDF:

\begin{align}
\label{mut-info}
R(\bf{D})=\min_{\bf{u}} \it{I}\left(\bf{y};\bf{u}|\bf{z}\right) 
\,\,\,  \text{         s.t.        }  \,\,\, \left\{ \!\!\! \begin{array}{ll} 
{{\bf{\Sigma }}_{{\bf{x}}|{\bf{u}}{{\bf{z}}}}} \preceq \bf{D}, \\
{\bf{u}} \leftrightarrow {\bf{y}} \leftrightarrow ({\bf{x}},{\bf{z}}).
\end{array} \right.
\end{align} 

We write ${\bf{x}}$ in terms of its linear estimation from ${\bf{y}}$ and ${\bf{z}}$ and the estimation error ${\bf{n}}$ as follows:

\begin{equation}
\label{xyz}
\bf{x}=\bf{Ay+Bz}+\bf{n},
\end{equation}

\noindent
where ${\bf{n}}$ is independent of ${\bf{y}}$ and ${\bf{z}}$, and ${\bf{\Sigma}}_{{\bf{n}}}={{\bf{\Sigma }}_{{\bf{x}}|{{\bf{yz}}}}}$. See Appendix \ref{appendix2} for a quick derivation of the matrices ${\bf{A}}$ and ${\bf{B}}$. 

Defining ${\bf{y}}'$ as:

\begin{equation}
\label{yprime}
{\bf{y}}'=\bf{Ay},
\end{equation}

\noindent
and using (\ref{xyz}) and the Markov chain in (\ref{mut-info}), one can write:

\begin{align}
\label{concov1}
& {{\bf{\Sigma }}_{{\bf{x}}|{{\bf{z}}}}} = {{\bf{\Sigma }}_{{\bf{y}}'|{{\bf{z}}}}} + {{\bf{\Sigma }}_{{\bf{x}}|{{\bf{yz}}}}}, \\
\label{concov2}
& {{\bf{\Sigma }}_{{\bf{x}}|{{\bf{uz}}}}} = {{\bf{\Sigma }}_{{\bf{y}}'|{{\bf{uz}}}}} + {{\bf{\Sigma }}_{{\bf{x}}|{{\bf{yz}}}}}.
\end{align} 

We solve Problem (\ref{mut-info}) in the rest of this section. To do so, we first introduce a notion of minimum for two positive-definite matrices in the next subsection.

%********************************************************************************************************************************************************
\subsection{Minimum of Two Matrices}
\label{diag}

Based on the joint diagonalization of two matrices, we define a minimum for two positive-definite matrices and derive some important properties of this definition. These properties will be crucial in the derivation of the RDF in the next two subsections. Although any joint diagonalization would work for this matter, we focus on a particular case which is defined in the sequel.

Consider the ordered pair $\left( {\bf{\Sigma}}_1 , {\bf{\Sigma}}_2 \right)$ of two $n \times n$ symmetric positive-definite matrices. Denote the eigenvalue decomposition of ${\bf{\Sigma}}_1$ by:

\begin{equation*}
{\bf{\Sigma}}_1 = {\bf{U}}^T \bf{\Lambda U},
\end{equation*}

\noindent
where ${\bf{\Lambda}}=\text{diag}\{\lambda_i, i=1,\dots,n\}$ and $\lambda_1 \geq \lambda_2 \geq \dots \geq \lambda_n$. Consider the principal square-root ${\bf{\Sigma}}_1^{\frac{1}{2}}= {\bf{U}}^T {\bf{\Lambda}}^{{\frac{1}{2}}} \bf{U}$ of ${\bf{\Sigma}}_1$ and denote the eigenvalue decomposition of ${\bf{\Sigma}}_1^{-{\frac{1}{2}}} {\bf{\Sigma}}_2 {\bf{\Sigma}}_1^{-{\frac{1}{2}}}$ by:

\begin{equation*}
{\bf{\Sigma}}_1^{-{\frac{1}{2}}} {\bf{\Sigma}}_2 {\bf{\Sigma}}_1^{-{\frac{1}{2}}} = {\bf{W}}^T \bf{\Gamma W},
\end{equation*}

\noindent
such that ${\bf{\Gamma}}=\text{diag}\{\gamma_i, i=1,\dotsc,n\}$ and $\gamma_1 \geq \gamma_2 \geq \dots \geq \gamma_n$. We then define the joint diagonalizer of $\left( {\bf{\Sigma}}_1 , {\bf{\Sigma}}_2 \right)$ as follows.

\begin{definition}
\label{defn_joint_diag}
The joint diagonalizer ${\bf{V}}$ of the ordered pair $\left( {\bf{\Sigma}}_1 , {\bf{\Sigma}}_2 \right)$ is defined as:

\begin{equation*}
{\bf{V}} = {\bf{\Lambda}}^{{\frac{1}{2}}} {\bf{W}} {\bf{\Sigma}}_1^{-{\frac{1}{2}}}.
\end{equation*}

\end{definition}

One can verify that:

\begin{align*}
&{\bf{V}} {\bf{\Sigma}}_1 {\bf{V}}^T = {\bf{\Lambda}}, \\
&{\bf{V}} {\bf{\Sigma}}_2 {\bf{V}}^T = {\bf{\Lambda}}',
\end{align*}

\noindent
where the diagonal matrix ${\bf{\Lambda}}' = \text{diag}\{\lambda'_i, i=1,\dotsc,n\}$ is defined as ${\bf{\Lambda}}' = {\bf{\Lambda}} {\bf{\Gamma}}$. 

This variant of joint diagonalization is based on another form of diagonalization found in \cite[Theorem 8.3.1]{matrix}. Based on the above joint diagonalization, we define the minimum of $\left( {\bf{\Sigma}}_1 , {\bf{\Sigma}}_2 \right)$ as follows:

\begin{definition}
The minimum of the pair $\left( {\bf{\Sigma}}_1 , {\bf{\Sigma}}_2 \right)$ of symmetric positive-definite matrices with the joint diagonalizer ${\bf{V}}$ is defined as:
\end{definition}

\vspace{-4 mm}
\begin{equation*}
{\min \left( {\bf{\Sigma}}_1 , {\bf{\Sigma}}_2 \right)} = {{\bf{V}}^{-1}} {\text{diag}} \{\min(\lambda_i,\lambda'_i), i=1,\dotsc,n\} {\bf{V}}^{-T}.
\end{equation*}

Some properties of the above definitions are summarized in the following lemmas.

\begin{lemma}
\label{lem_properties}
For the pair $\left( {\bf{\Sigma}}_1 , {\bf{\Sigma}}_2 \right)$ of symmetric positive-definite matrices with the joint diagonalizer ${\bf{V}}$ the following properties hold:

\begin{enumerate}
\item $|{\bf{V}}| = 1$.
\item $\lambda'_1 \geq \lambda'_2 \geq \dots \geq \lambda'_n$.
\item ${\min{ \left( {\bf{\Sigma}}_1 , {\bf{\Sigma}}_2 \right)}} \preceq {\bf{\Sigma}}_1$, and ${\min \left( {\bf{\Sigma}}_1 , {\bf{\Sigma}}_2 \right)} \preceq {\bf{\Sigma}}_2$.
\item ${\min \left( {\bf{\Sigma}}_1 , {\bf{\Sigma}}_2 \right)} = {\bf{\Sigma}}_1$ if and only if ${\bf{\Sigma}}_1 \preceq {\bf{\Sigma}}_2$, and ${\min \left( {\bf{\Sigma}}_1 , {\bf{\Sigma}}_2 \right)} = {\bf{\Sigma}}_2$ if and only if ${\bf{\Sigma}}_2 \preceq {\bf{\Sigma}}_1$.
\end{enumerate}
\end{lemma}

\begin{IEEEproof}
The proof follows immediately from the definitions and is thus left out.
\end{IEEEproof}

The proof of the following lemma follows from the results in \cite{Li}.

\begin{lemma}
\label{lem_properties2}
Consider the following optimization problem:

\begin{align*}
%\label{}
 a^* = \max_{ {\bf{\Sigma}} \in {\mathbb{S}}^+} | {\bf{\Sigma}} | 
\,\,\, \text{ s.t. } \left\{ \!\!\! \begin{array}{ll} 
{\bf{\Sigma}} \preceq {\bf{\Sigma}}_1, \\
{\bf{\Sigma}} \preceq {\bf{\Sigma}}_2,
\end{array} \right.
\end{align*}

\noindent
where ${\bf{\Sigma}}, {\bf{\Sigma}}_1, {\bf{\Sigma}}_2 \in {\mathbb{S}}^+$. We then have $a^* = | {\min \left( {\bf{\Sigma}}_1 , {\bf{\Sigma}}_2 \right)} |$.

\end{lemma}

%\begin{IEEEproof}
%See Appendix \ref{appendix1}.
%\end{IEEEproof}

%********************************************************************************************************************************************************
\subsection{Rate Distortion Function}
\label{theRDF}

The following theorem is the main result of this section.

\begin{theorem}
\label{th1}
The rate-distortion function for problem (\ref{mut-info}) for ${\bf{D}} \succ {{\bf{\Sigma }}_{{\bf{x}}|{\bf{y}}{{\bf{z}}}}}$ is given by:

\begin{equation}
\label{finalRDF}
R({\bf{D}}) = \frac{1}{2} \log \frac{\left| {{\bf{\Sigma }}_{\bf{x|z}}} - {{\bf{\Sigma }}_{\bf{x|yz}}} \right|}{\left| \min ({\bf{D}} - {{\bf{\Sigma }}_{\bf{x|yz}}} , {{\bf{\Sigma }}_{\bf{x|z}}} - {{\bf{\Sigma }}_{\bf{x|yz}}}) \right|}.
\end{equation}

\end{theorem}

\begin{IEEEproof}
We prove the theorem by deriving coinciding upper and lower bounds on the RDF in Sections \ref{lower} and \ref{upper}, respectively.
\end{IEEEproof}

%********************************************************************************************************************************************************
\subsection{Lower Bound on $R(\bf{D})$}
\label{lower}

Let $\bf{V}$ be the joint diagonalizer of $( {{\bf{\Sigma }}_{\bf{x|z}}} - {{\bf{\Sigma }}_{\bf{x|yz}}} , {\bf{D}} - {{\bf{\Sigma }}_{\bf{x|yz}}} )$ as given in Definition \!\ref{defn_joint_diag}, such that: \!\footnote{We assume that ${{\bf{\Sigma }}_{\bf{x|z}}} - {{\bf{\Sigma }}_{\bf{x|yz}}}$ is of full rank. The rank-deficient case can be converted to an equivalent problem with full-rank matrices using an appropriate transformation. See e.g. Appendix A in \cite{Wagner2012}.}

\begin{align}
\label{joint1}
&{\bf{V}} ({{\bf{\Sigma }}_{\bf{x|z}}} - {{\bf{\Sigma }}_{\bf{x|yz}}}) {\bf{V}}^T = {\bf{\Lambda}}, \\
\label{joint2}
&{\bf{V}} ({\bf{D}} - {{\bf{\Sigma }}_{\bf{x|yz}}}) {\bf{V}}^T = {\bf{\Lambda}}'.
\end{align}

\noindent
Define the function $\tilde{R} ({\bf{D}})$ as:

\begin{equation}
\label{explicit}
\tilde{R} ({\bf{D}}) = \frac{1}{2} \log \frac{\left| {{\bf{\Sigma }}_{\bf{x|z}}} - {{\bf{\Sigma }}_{\bf{x|yz}}} \right|}{\left| \min ({\bf{D}} - {{\bf{\Sigma }}_{\bf{x|yz}}} , {{\bf{\Sigma }}_{\bf{x|z}}} - {{\bf{\Sigma }}_{\bf{x|yz}}}) \right|}.
\end{equation}

\noindent
The following lemma is then the main result of this subsection.

\begin{lemma}
\label{lem1}
The RDF $R({\bf{D}})$ defined in (\ref{mut-info}) is lower-bounded as $R({\bf{D}}) \geq \tilde{R} ({\bf{D}})$.
\end{lemma}

\begin{IEEEproof}
See Appendix \ref{appendix_lowerbound}.
\end{IEEEproof}

%\vspace{-4mm}
%********************************************************************************************************************************************************
\subsection{Upper Bound on $R(\bf{D})$}
\label{upper}

In this subsection, we upper bound $R(\bf{D})$ with the same function $\tilde{R} ({\bf{D}})$ which appeared in the lower bound in the previous subsection. This thus in combination with the lower-bound, gives the RDF in a closed form. It is important to note though that the knowledge of the RDF does not necessarily specify how one could encode the observations to achieve $R(\bf{D})$. The upper bound derived in this subsection is constructive, in the sense that it is based on an achievable coding scheme. Such a scheme should suggest a way to transform ${\bf{y}}$ into another variable ${\bf{u}}$, such that for a given distortion ${\bf{D}}$:

\begin{itemize}
\item The required rate for delivering ${\bf{u}}$ to the decoder is no higher than $R(\bf{D})$.
\item The decoder could make an estimate ${\bf{\hat{x}}}$ of ${\bf{x}}$ using ${\bf{u}}$ and ${\bf{z}}$, such that the distortion constraint (\ref{covariance_distortion_constraint}) is satisfied.
\end{itemize}

The role of ${\bf{u}}$ here is to model the quantization effect on ${\bf{y}}$ (or a transformation of ${\bf{y}}$). It is thus typically composed of a linear transformation of ${\bf{y}}$ and an additive noise term that represents the quantization noise. We suggest such a scheme in the sequel. Suppose that $\bf{V}$, ${\bf{\Lambda}} = {\text{diag}} \{ \lambda_1,\cdots,\lambda_n \} $, and ${\bf{\Lambda}}' = {\text{diag}} \{ \lambda'_1,\cdots,\lambda'_n \}$ are defined as in (\ref{joint1}) and (\ref{joint2}). Suppose also that $\bf{U}$ is the orthogonal matrix of the eigenvectors of ${{\bf{\Sigma }}_{\bf{x|z}}} - {{\bf{\Sigma }}_{\bf{x|yz}}}$ giving the eigenmatrix ${\bf{\Lambda}}$. The result of this subsection is then given by the following lemma.

\begin{lemma}
\label{lem3}
The following Gaussian test channel achieves $\tilde{R} ({\bf{D}})$ given by (\ref{explicit}):

\begin{equation}
\label{scheme}
{\bf{u}}^* = {\bf{UAy}} + {\boldsymbol{\nu}},
\end{equation}

\noindent
where ${\bf{A}}$ is defined in (\ref{xyz}), and the the coding noise $\boldsymbol{\nu}$ is independent of ${\bf{y}}$ with the covariance matrix given by:

\begin{equation}
\label{cov_noise}
{{\bf{\Sigma }}_{\bf{\nu}}} = {{\bf{UV}}^{-1}}{\rm{diag}}\left\{\! { \frac{{\lambda _i}\min \left({\lambda _i},{\lambda' _i}\right)}{{\lambda _i} \!-\! \min \left({\lambda _i},{\lambda' _i} \right)} } , i=1,\dotsc,n_x \! \right\}\!{{{\bf{V}}^{-T}}{{\bf{U}}^{T}}}.
\end{equation}

\noindent
Moreover, the covariance matrix of the reconstruction error at the optimal decoder is given by:

\begin{equation}
\label{rec_err_formula}
{{\bf{\Sigma }}_{{\bf{e}}}} = {{\bf{\Sigma }}_{{\bf{x}}|{\bf{u^*}}{{\bf{z}}}}} = {{\bf{\Sigma }}_{{\bf{x}}|{\bf{y}}{{\bf{z}}}}} + {\min \left( {{\bf{\Sigma }}_{\bf{x|z}}} \!-\! {{\bf{\Sigma }}_{\bf{x|yz}}} , {\bf{D}} \!-\! {{\bf{\Sigma }}_{\bf{x|yz}}} \right)}.
\end{equation}

\end{lemma}

\begin{IEEEproof}
See Appendix \ref{appendix_upperbound}.
\end{IEEEproof}

\vspace{6mm}
Note that the coding scheme given by (\ref{scheme}) involves two steps:

\begin{itemize}
\item Linear transformation of the observation ${\bf{y}}$ into ${\bf{UAy}}$,
\item Quantization of the resulting sequence such that the quantization noise ${\boldsymbol{\nu}}$ becomes Gaussian and independent of ${\bf{y}}$ with the covariance matrix given by (\ref{cov_noise}).
\end{itemize}

\noindent
The second step can be performed using a high-dimensional dithered vector quantizer. The dithered quantizer guarantees that ${\boldsymbol{\nu}}$ will be independent of ${\bf{y}}$, and the high dimension allows for the Gaussianity of ${\boldsymbol{\nu}}$ \cite{ram}.

Finally, notice that an immediate result of the above lemma is that $R({\bf{D}}) \leq \tilde{R} ({\bf{D}})$, which establishes the desired upper bound on the RDF.

\emph{Remark:} The first and second terms at the right-hand side of (\ref{rec_err_formula}) are related to the error due to the remoteness of the source and the coding artifacts, respectively. This separability resembles the Wolf and Ziv's result in \cite{Jack-Wolf}, where it is shown that for a point-to-point remote joint source-channel coding problem with mean-squared error distortions, the end-to-end distortion can be decomposed as the sum of two terms. One term is due to the remoteness of the source and the other one is due to coding. Note however, that it was shown in \cite{Relevant}, that this result is not optimal for the multiterminal case.

%********************************************************************************************************************************************************
%********************************************************************************************************************************************************
%********************************************************************************************************************************************************
%********************************************************************************************************************************************************

\section{Applications and Special Cases}
\label{applications}

%********************************************************************************************************************************************************

\subsection{Mean-Squared Error Constraint}
\label{mse}
For the vector Gaussian remote Wyner-Ziv problem with a scalar mean-squared error constraint, the rate-distortion function is given by \cite{Tian}:

\begin{align*}
%\label{rdf_mse}
R(D) \!=\! \min_{\bf{u}} \!I\! \left(\bf{y};\bf{u}|\bf{z}\right) \,\,\,\, \text{s.t.   } \,\,\, \text{tr}\left( {{\bf{\Sigma }}_{{\bf{x}}|{\bf{u}}{{\bf{z}}}}} \right) \leq n_x D, {\bf{u}} \!\leftrightarrow\! {\bf{y}} \!\leftrightarrow\! ({\bf{x}},{\bf{z}}).
\end{align*} 

It was shown in \cite{Tian} that $R(D)$ is given by:

\begin{equation*}
%\label{rdf_mse2}
R(D)= \frac{1}{2}\sum_{i=1}^{n_x} {\log {{\left( {\frac{{{\lambda _i}}}{\lambda }} \right)}^+ }} ,
\end{equation*}

\noindent
where $\lambda _i; \: i = 1,2,\dotsc,n_x$ are the eigenvalues of ${{\bf{\Sigma }}_{\bf{x|z}}} - {{\bf{\Sigma }}_{\bf{x|yz}}}$, and $\lambda$ is a constant (water level) satisfying the following constraint:

\begin{equation}
\label{water}
\sum\limits_{i = 1}^{{n_x}} {\min \left( {\lambda ,{\lambda _i}} \right)}  = {n_x}D - {\rm{tr}}\left( {{\bf{\Sigma} _{\bf{x|yz}}}} \right).
\end{equation}

We will show that this is equivalent to a special case of $R(\bf{D})$ for a particular choice of the distortion matrix $\bf{D}$. Denote the eigenvalue decomposition of ${{\bf{\Sigma }}_{{\bf{x}}|{{\bf{z}}}}} - {{\bf{\Sigma }}_{\bf{x|yz}}}$ by ${\bf{U}}^T \text{diag}\{\lambda_i, i=1,\dotsc,n_x\} \bf{U}$, and define:

\begin{equation*}
%\label{opt_D_mse}
{\bf{D}}^* = {{\bf{\Sigma} _{\bf{x|yz}}}} + {\bf{U}}^T \text{diag}\{\min(\lambda,\lambda_i), i=1,\dotsc,n_x\} \bf{U}, 
\end{equation*}

\noindent 
where $\lambda$ is defined by (\ref{water}). It is clear that ${\bf{D}}^* \succ {{\bf{\Sigma} _{\bf{x|yz}}}}$ and is thus a valid distortion matrix. Substituting ${\bf{D}}^*$ in (\ref{finalRDF}) yields:

\begin{align*}
%\label{UUBB}
R({\bf{D}}^*) &  =  \frac{1}{2}\sum_{i=1}^{n_x} {\log {{\left( {\frac{{{\lambda _i}}}{ \min{(\lambda,\lambda_i)} }} \right)} }}  \nonumber \\
& = \frac{1}{2}\sum_{i=1}^{n_x} {\log {{\left( {\frac{{{\lambda _i}}}{\lambda }} \right)}^+ }} = R(D).
\end{align*}

%********************************************************************************************************************************************************

\subsection{Relay Networks}
\label{mutual_info}

Consider a relay network where a main transmitter uses $n_x$ antennas to transmit the Gaussian signal ${\bf{x}}$ to the end receiver which has its own noisy observation ${\bf{z}}$ of ${\bf{x}}$ using $n_z$ antennas. There is also a relay that makes the noisy observation ${\bf{y}}$ using its $n_y$ antennas and transmits to the end receiver with a given maximum rate. It is desired to find the minimum rate at which the system can reliably transmit. Assuming that the statistics and channel state information for ${\bf{x}}$ are known to the relay and receiver, it was shown in \cite{Tian} that this problem is equivalent to the following optimization problem:

\begin{align*}
%\label{rif}
R(R_I)=\min_{\bf{u}} I\left(\bf{y};\bf{u}|\bf{z}\right) \,\,\,\, \text{s.t.} \,\,\,\, I\! \left( {{\bf{x}};{\bf{u}}|{{\bf{z}}}} \right) \!\geq\! R_I, {\bf{u}} \!\leftrightarrow\! {\bf{y}} \!\leftrightarrow\! ({\bf{x}},{\bf{z}}),
\end{align*}

\noindent
for a given $R_I$, and the resulting rate-mutual information function $R(R_I)$ for $0 \leq R_I \leq \frac{1}{2}\log \left( {\frac{{\left| {{{\bf{\Sigma }}_{{\bf{x}}|{{\bf{z}}}}}} \right|}}{{\left| { {{\bf{\Sigma }}_{{\bf{x}}|{\bf{y}}{{\bf{z}}}}}} \right|}}} \right)$  is given by:

\begin{equation}
\label{rate-rate}
{R}\left( {R_I} \right) = \frac{1}{2}\sum_{i=1}^{n_x} { \log \left( {\mu _i} {\left[ {\left( \frac{1 - {\mu _i}}{1 - \gamma} \right)}^{-} - \left( 1 - {\mu _i} \right) \right] }^ {-1} \right) },
\end{equation}

\noindent
where ${\mu _i}$, $i=1,\dotsc,n_x$ are the eigenvalues of ${\bf{I}}_{n_x} - {{\bf{\Sigma }}^{-\frac{1}{2}}_{\bf{x|z}}} {{\bf{\Sigma }}_{\bf{x|yz}}} {{\bf{\Sigma }}^{-\frac{1}{2}}_{\bf{x|z}}}$, \footnote{In \cite{Tian}, ${\mu _i}$ are defined as the nonzero eigenvalues of ${{\bf{\Sigma}}_{{\bf{y|z}}}^{1/2}} {{\bf{A}}^T} {{\bf{\Sigma}}_{{\bf{x|z}}}^{-1}} {\bf{A}} {{\bf{\Sigma}}_{{\bf{y|z}}}^{1/2}}$. Note however that for any matrix ${\bf{Q}}$ the zero-excluded multispectrum is the same for ${\bf{Q}} {\bf{Q}}^T$ and ${\bf{Q}}^T {\bf{Q}}$. One can then see that ${\mu _i}$ are also the eigenvalues of ${{\bf{\Sigma}}_{{\bf{x|z}}}^{-1/2}} {{\bf{A}}} {{\bf{\Sigma}}_{{\bf{y|z}}}} {\bf{A}}^T {{\bf{\Sigma}}_{{\bf{x|z}}}^{-1/2}}$. Using (\ref{yprime}) and (\ref{concov1}), this simplifies to ${\bf{I}}_{n_x} - {{\bf{\Sigma }}^{-\frac{1}{2}}_{\bf{x|z}}} {{\bf{\Sigma }}_{\bf{x|yz}}} {{\bf{\Sigma }}^{-\frac{1}{2}}_{\bf{x|z}}}$.} and $\gamma \in [0,1)$ satisfies the following:

\begin{eqnarray}
\label{mut-inf-water}
{-\frac{1}{2}} \sum_{i=1}^{n_x} {\log {\left( \frac{1 - {\mu _i}}{1 - \gamma} \right)}^{-} } = R_I.
\end{eqnarray}

\noindent See \cite{Tian} and the references therein for more details on the problem setup and formulation. We will first show that ${R}\left( {R_I} \right)$ is a special case of $R({\bf{D}})$ for a particular choice of the distortion matrix ${\bf{D}}$, and then using the results of Section \ref{WZ}, we will show how to design a coding scheme that achieves $R(R_I)$.

Suppose that the eigenvalue decomposition of ${\bf{I}}_{n_x} - {{\bf{\Sigma }}^{-\frac{1}{2}}_{\bf{x|z}}} {{\bf{\Sigma }}_{\bf{x|yz}}} {{\bf{\Sigma }}^{-\frac{1}{2}}_{\bf{x|z}}}$ is given by:

\begin{equation}
\label{new_eig}
{\bf{I}}_{n_x} - {{\bf{\Sigma }}^{-\frac{1}{2}}_{\bf{x|z}}} {{\bf{\Sigma }}_{\bf{x|yz}}} {{\bf{\Sigma }}^{-\frac{1}{2}}_{\bf{x|z}}} = {\bf{W}}^T \text{diag}\{\mu_i, i=1,\dotsc,n_x\} \bf{W}.
\end{equation}

\noindent
Define:

\begin{equation}
\label{opt_D_inf}
{\bf{D}}^* \!=\! {{\bf{\Sigma }}^{\frac{1}{2}}_{\bf{x|z}}}{\bf{W}}^T \text{diag} \! \left\{ \! \min \! \left(\! 1,\frac{1-\mu_i}{1-\gamma} \! \right), i=1,\dotsc,n_x \!\right\}\! {\bf{W}} {{\bf{\Sigma }}^{\frac{1}{2}}_{\bf{x|z}}},
\end{equation}

\noindent
where $\gamma$ is given by (\ref{mut-inf-water}). From the facts that $\gamma < 1$ and ${{\bf{\Sigma }}^{-\frac{1}{2}}_{\bf{x|z}}} {{\bf{\Sigma }}_{\bf{x|yz}}} {{\bf{\Sigma }}^{-\frac{1}{2}}_{\bf{x|z}}} \preceq {\bf{I}}_{n_x}$, and using (\ref{new_eig}) and (\ref{opt_D_inf}), it follows that ${{\bf{\Sigma }}^{-\frac{1}{2}}_{\bf{x|z}}} {\bf{D}}^* {{\bf{\Sigma }}^{-\frac{1}{2}}_{\bf{x|z}}} \succ {{\bf{\Sigma }}^{-\frac{1}{2}}_{\bf{x|z}}}{{\bf{\Sigma }}_{\bf{x|yz}}} {{\bf{\Sigma }}^{-\frac{1}{2}}_{\bf{x|z}}}$, or equivalently  ${\bf{D}}^* \succ {{\bf{\Sigma }}_{\bf{x|yz}}}$. Therefore, ${\bf{D}}^*$ is a valid distortion matrix. Substituting ${\bf{D}}^*$ in (\ref{finalRDF}), noting that ${\bf{D}}^* \preceq {{\bf{\Sigma }}_{\bf{x|z}}}$, and using property 4 in Lemma \ref{lem_properties}, we have:

\begin{align}
R({\bf{D}}^*)&  = \frac{1}{2} \log \frac{\left| {{\bf{\Sigma }}_{\bf{x|z}}} - {{\bf{\Sigma }}_{\bf{x|yz}}} \right|}{\left| \min ({\bf{D}}^* - {{\bf{\Sigma }}_{\bf{x|yz}}} , {{\bf{\Sigma }}_{\bf{x|z}}} - {{\bf{\Sigma }}_{\bf{x|yz}}}) \right|} \nonumber \\
&  = \frac{1}{2} \log \frac{\left| {{\bf{\Sigma }}_{\bf{x|z}}} - {{\bf{\Sigma }}_{\bf{x|yz}}} \right|}{\left| {\bf{D}}^* - {{\bf{\Sigma }}_{\bf{x|yz}}} \right|} \nonumber \\
&=\frac{1}{2} \log \frac{\left| {\bf{I}}_{n_x} - {{\bf{\Sigma }}^{-\frac{1}{2}}_{\bf{x|z}}} {{\bf{\Sigma }}_{\bf{x|yz}}} {{\bf{\Sigma }}^{-\frac{1}{2}}_{\bf{x|z}}} \right|}{\left| {{\bf{\Sigma }}^{-\frac{1}{2}}_{\bf{x|z}}} {\bf{D}}^* {{\bf{\Sigma }}^{-\frac{1}{2}}_{\bf{x|z}}} - {{\bf{\Sigma }}^{-\frac{1}{2}}_{\bf{x|z}}} {{\bf{\Sigma }}_{\bf{x|yz}}} {{\bf{\Sigma }}^{-\frac{1}{2}}_{\bf{x|z}}} \right|} \nonumber \\
\label{UB_opt3}
& = \frac{1}{2} \sum\limits_{i=1}^{n_x} \log \left( \frac{\mu_i}{1-\mu_i} \frac{1-\gamma}{\gamma} \right)^+  \\
\label{UB_opt2}
& = R(R_I).
\end{align}

\noindent
where (\ref{UB_opt2}) follows from some straightforward manipulation of (\ref{rate-rate}) and comparing the result with (\ref{UB_opt3}). Recall that for any valid distortion matrix ${\bf{D}}$, the coding scheme (\ref{scheme}) achieves $R({\bf{D}})$. Since ${\bf{D}}^*$ given by (\ref{opt_D_inf}) is a valid distortion matrix, to implement a coding scheme that yields $R(R_I)= R({\bf{D}}^*)$, one could set ${\bf{D}} = {\bf{D}}^*$ and use (\ref{scheme}).

%********************************************************************************************************************************************************

\subsection{Noise Reduction in Sensor Networks}
\label{enhance}

\begin{figure*}[tb]
\begin{center}
\includegraphics[scale=0.35]{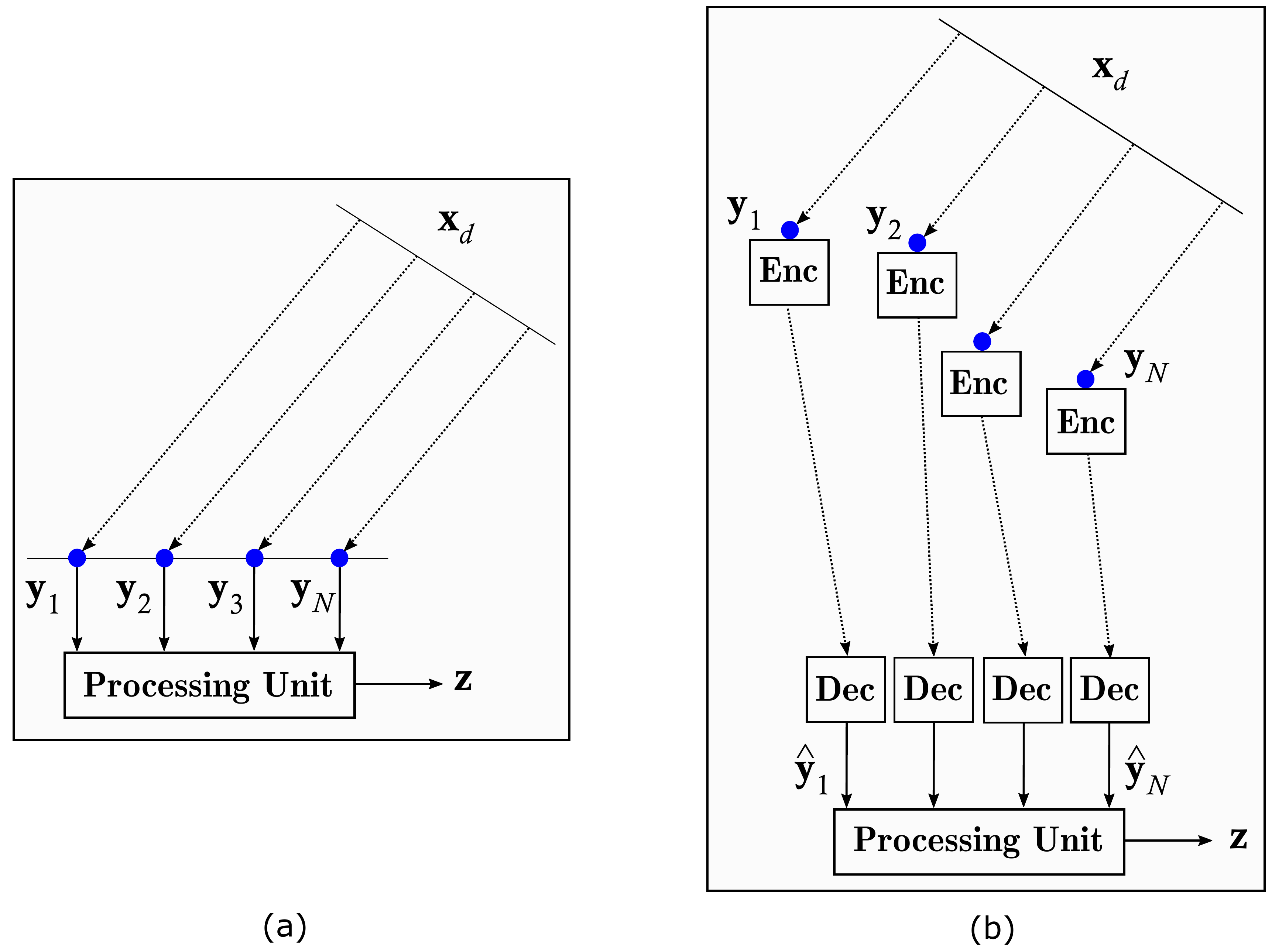}
\caption{(a) A sensor array, (b) a centralized wireless sensor network.}
\label{array_network}
\end{center}
\vspace{-2 mm}
\end{figure*}

A simple diagram of a sensor array as well as a centralized network of wireless sensors is shown in Fig. \ref{array_network}. In both cases, there is a central processing unit which makes an estimation of the desired source ${\bf{x}}_d$ using the received signals. For the sensor array, the signals which are fed to the processing unit are impaired versions of the desired source affected by the environment. One can then design the processing unit so as to minimize the noise at the output. For the sensor network, the sensor signals which are already impaired by the environment effects, have to be transmitted to the center with limited rates via wireless links. This incurs extra noise terms which are due to digital transmission (coding noise). Although this extra noise term degrades the performance, its impact can be alleviated by the fact that the coding noise can be controlled by the designer. In other words, in addition to the processing unit, one can make use of the distortion matrices as design parameters in order to minimize the output noise at the processing center. Let us elaborate on this idea in the sequel.

%\vspace{1cm}

%\vspace{-4mm}

Consider an array of $N$ sensors, where sensor $i,\; i=1,\cdots,N$ observes a noisy version $y_i(k)$ of a Gaussian source $x_i(k)$ given by:

\begin{equation}
\label{scalar_samples}
y_i(k) = x_i(k) + n_i(k),
\end{equation}

\noindent
where $k$ is the time index, and $n_i(k)$ is the additive noise which is assumed to be Gaussian and independent of the source $x_i(k)$ and the noise terms at the other nodes. Define ${\bf{y}}_i \in {\mathbb{R}}^n$ as ${\bf{y}}_i = [y_i(k-n+1) \; \cdots \; y_i(k)]^T$ (and similarly for ${\bf{x}}_i$ and ${\bf{n}}_i$). One could then rewrite (\ref{scalar_samples}) in the vector form as:

\begin{equation}
\label{vector_samples}
{\bf{y}}_i = {\bf{x}}_i + {\bf{n}}_i,
\end{equation}

\noindent
where we drop the dependency on $k$ for simplicity of notation. The sources ${\bf{x}}_i; i=1,\cdots,N$ are related to a desired source ${\bf{x}}_d$ as:

\begin{equation}
\label{predictablity}
{\bf{x}}_i = {\bf{W}}^T_i \, {\bf{x}}_d,  \; i=1,...,N,
\end{equation}

\noindent
where ${\bf{W}}_i, \; i=1,\cdots,N$ are $n \times n$ matrices. Defining ${\bf{W}}$ as ${\bf{W}} = [{\bf{W}}_1 \; ... \; {\bf{W}}_N]$, we rewrite (\ref{predictablity}) as:

\begin{equation*}
%\label{predictablity2}
{\bf{x}} = {\bf{W}}^T {\bf{x}}_d.
\end{equation*}

\noindent
where ${\bf{x}} = [{\bf{x}}_1^T \; ... \; {\bf{x}}_N^T]^T$ (and we also define $\bf{y}$ and $\bf{n}$ similarly). At the processing unit, as shown in Fig. \!\ref{array_network} (a), the output ${\bf{z}}$ is produced using the following filtering operation:

\begin{equation*}
%\label{filtering}
{\bf{z}} = \sum_{i=1}^N {\bf{H}}_i {\bf{y}}_i = {\bf{Hy}} = {\bf{Hx}} + {\bf{Hn}},
\end{equation*}

\noindent
where ${\bf{H}} = [{\bf{H}}_1 \; ... \; {\bf{H}}_N]$. The objective is to maximize the output SNR of the filter ${\bf{H}}$ while enforcing only additive distortion to the reconstruction (no linear distortion). A typical example of this scenario with no linear distortion is in microphone arrays, where the linear distortion would cause undesired artefacts in the resulting audio or speech signal \cite{benesty}. Another example of applications is in Networked Control Systems, where the feedback data is transmitted via a communication network. In certain cases, because of the implementation constraints the source coding scheme used to compress the feedback information should incur no linear distortion \cite{control}.

To formulate the problem, notice that the output error of the filter ${\bf{H}}$ is given by:

\begin{equation*}
%\label{error}
{\bf{e}} = {\bf{z}} - {\bf{x}}_d = \left( {\bf{HW}}^T - {\bf{I}} \right) {\bf{x}}_d + {\bf{Hn}} = {\bf{e_x}} + {\bf{e_n}},
\end{equation*}

\noindent
where ${\bf{e_x}}$ is the linear distortion of the signal and ${\bf{e_n}}$ is the additive distortion. To minimize the additive distortion while suppressing the linear distortion, one could choose ${\bf{H}} = {\bf{H}}_{nld}$, where

\begin{equation}
\label{Hst}
{\bf{H}}_{nld} = \text{arg}\min_{\bf{H}} \text{tr}(E[{\bf{e_n}} {\bf{e_n}}^T]) \;\;\;\; s.t. \;\;\;\; {\bf{HW}}^T = {\bf{I}}.
\end{equation}

There is a closed form solution for (\ref{Hst}) given by \cite{benesty}:

\begin{equation}
\label{Best-filter}
{\bf{H}}_{nld} = \left( {\bf{W}} {\bf{\Sigma}}_{\bf{n}}^{-1} {\bf{W}}^T \right)^{-1} {\bf{W}} {\bf{\Sigma}}_{\bf{n}}^{-1},
\end{equation}

\noindent
which gives the following SNR at the output:

\begin{equation}
\label{SNR}
\text{SNR}({\bf{H}}_{nld}) = \frac{\text{tr}\left( {\bf{\Sigma}}_{{\bf{x}}_d} \right)}{\text{tr}\left\{ \left( {\bf{W}} {\bf{\Sigma}}_{\bf{n}}^{-1} {\bf{W}}^T \right)^{-1} \right\}}.
\end{equation}

Now suppose that instead of $N$ array sensors, we have a network of $N$ wireless sensors as shown in Fig. \!\ref{array_network} (b). Node $i $ encodes its observation ${\bf{y}}_i$ and transmits it with the rate $R({\bf{D}}_i)$ for a target distortion ${\bf{D}}_i$. This RDF is equivalent to (\ref{finalRDF}) for the special case where $n_x = n_y = n_z = n$, ${\bf{x}} = {\bf{y}}$ and ${\bf{z}} = {\bf{0}}$. The result is given by:

\begin{equation*}
%\label{RDF_enhance1}
R({\bf{D}}_i) = \frac{1}{2} \log \frac{\left| {{\bf{\Sigma }}_{{\bf{y}}_i}} \right|}{\left| \min ({\bf{D}}_i , {{\bf{\Sigma }}_{{\bf{y}}_i}} ) \right|}.
\end{equation*}

\noindent
We further assume that ${\bf{D}}_i \preceq {{\bf{\Sigma }}_{{\bf{y}}_i}}$. From property 4 in Lemma \ref{lem_properties} it then follows that:

\begin{equation}
\label{RDF_enhance2}
R({\bf{D}}_i) = \frac{1}{2} \log \frac{\left| {{\bf{\Sigma }}_{{\bf{y}}_i}} \right|}{\left| {\bf{D}}_i \right|}.
\end{equation}

\noindent Note that we have assumed independent decoding of the messages received from the sensors. This is in general suboptimal. An optimal strategy requires a joint decoding similar to the CEO problem \cite{CEO}, which would be difficult to solve analytically and also to implement.

From the definition of the distortion constraint (\ref{covariance_distortion_constraint}) and the Gaussianity of the achievable coding scheme (see also the proof of Lemma \ref{lem3}), it follows that the reconstruction $\hat{\bf{y}}'_i$ at the decoder satisfies ${\bf{y}}_i - \hat{\bf{y}}'_i = {\boldsymbol{\nu}_i}$, with ${\boldsymbol{\nu}_i}$ being independent of $\hat{\bf{y}}'_i$ and ${\bf{\Sigma }}_{{\boldsymbol{\nu}_i}} = {\bf{D}}_i$. Equivalently, one could write:

\begin{equation*}
%\label{reconst}
\hat{\bf{y}}'_i =  {\bf{A}}_i {\bf{y}}_i + {\boldsymbol{\nu}'_i},
\end{equation*} 

\noindent
where ${\boldsymbol{\nu}'_i}$ and ${\bf{y}}_i$ are independent, and:

\begin{align*}
%\label{reconst2}
& {\bf{A}}_i = ({\bf{\Sigma}}_{{\bf{y}}_i} - {\bf{D}}_i) {\bf{\Sigma}}_{{\bf{y}}_i}^{-1} \\
%\label{reconst3}
& {\bf{\Sigma }}_{{\boldsymbol{\nu}'_i}} = {\bf{A}}_i {\bf{D}}_i.
\end{align*}

\noindent
We apply the invertible map ${\bf{A}}_i^{-1}$ to $\hat{\bf{y}}_i'$ to obtain the following equivalent (but preferred) version of the reconstruction:

\begin{equation}
\label{yhat1}
\hat{\bf{y}}_i = {\bf{A}}_i^{-1} \hat{\bf{y}}_i' = {\bf{y}}_i + {\boldsymbol{\nu}}_i,
\end{equation} 

\noindent
with:

\begin{equation}
\label{yhat2}
{\bf{\Sigma }}_{{\boldsymbol{\nu}}_i} = {\left( {\bf{D}}_i^{-1} - {\bf{\Sigma}}_{{\bf{y}}_i}^{-1} \right)}^{-1}.
\end{equation}

%\vspace{1cm}
%\begin{figure}[tb]
%\begin{center}
%\includegraphics[scale=0.25]{WASN}
%\caption{The wireless sensor network setup}
%\label{WASN}
%\end{center}
%\vspace{-2 mm}
%\end{figure}
%\vspace{1cm}

Combining (\ref{vector_samples}) and (\ref{yhat1}) we have:

\begin{equation}
\label{input1}
{\hat{\bf{y}}}_i = {\bf{x}}_i + {\bf{n}}_i + {\boldsymbol{\nu}}_i = {\bf{x}}_i + {\bf{v}}_i,
\end{equation}

\noindent
where ${\bf{v}}_i = {\bf{n}}_i + {\boldsymbol{\nu}}_i$. Recall that the objective is to maximize the SNR at the output of the processing unit using the received noisy signals ${\hat{\bf{y}}}_i$, while incurring no linear distortion to the output signal. Comparing (\ref{input1}) to (\ref{vector_samples}), it is clear that this problem is similar to the case of the sensor array. One can thus apply the same filter as (\ref{Best-filter}) to ${\hat{\bf{y}}}_i; i=1,\cdots,N$ to maximize the output SNR. To do so, we rewrite (\ref{input1}) as:

\begin{equation}
\label{input2}
{\hat{\bf{y}}} = {\bf{x}} + {\bf{v}},
\end{equation}

\noindent
where ${\hat{\bf{y}}} = [\hat{\bf{y}}_1^T \; ... \; \hat{\bf{y}}_N^T]^T$, and ${\bf{x}}$ and ${\bf{v}}$ are defined similarly. The output SNR similar to (\ref{SNR}) is then given by:

\begin{align}
\label{SNR2}
&\text{SNR}({\bf{H}}_{nld},{\bf{D}}_1, ..., {\bf{D}}_N) = \nonumber \\
&\frac{\text{tr}\left( {\bf{\Sigma}}_{{\bf{x}}_d} \right)}{\text{tr}\left\{ \left[ {\bf{W}} \left( {{\bf{\Sigma}}_{\bf{v}} ({\bf{D}}_1, ..., {\bf{D}}_N)} \right) ^{-1} {\bf{W}}^T \right]^{-1} \right\}}. 
\end{align}

Notice that due to the wireless link between the sensors and the center and the limited transmission rate, in addition to the additive noise ${\bf{n}}_i$ in the measurements, there is also a reconstruction error ${\boldsymbol{\nu}}_i$ due to quantization and coding. Therefore, the covariance matrix of the noise term ${\bf{v}}$ in (\ref{input2}) depends on the set of distortions as emphasized by the notation in (\ref{SNR2}). While the quantization error is inevitable, one could shape the covariance matrix ${{\bf{\Sigma}}_{\bf{v}} ({\bf{D}}_1, ..., {\bf{D}}_N)}$ in (\ref{SNR2}) by manipulating the distortions ${\bf{D}}_1, ..., {\bf{D}}_N$ in order to maximize the output SNR. 

To be more specific, suppose that it is required that the weighted sum-rate of the network $\sum_{i=1}^N \alpha_i R({\bf{D}}_i)$ does not exceed a given amount $R$ for a given set of weights $\alpha_1, ..., \alpha_N$. The problem can then be formulated as follows:

\begin{align}
\label{MIN}
\max_{{\bf{D}}_1, ..., {\bf{D}}_N} \text{SNR}({\bf{H}}_{nld},{\bf{D}}_1, ..., {\bf{D}}_N) \;\;\;\;  s.t. \;\;\;\;   \sum_{i=1}^N \alpha_i R({\bf{D}}_i) = R.
\end{align}

The weights $\alpha_1, ..., \alpha_N$ could be chosen for example to equalize the power consumption at the sensors, or to give the network sum-rate by being set equal to each other, or to satisfy any other criterion specified by the user. Without loss of generality, we assume that:

\begin{equation}
\label{weights}
\sum_{i=1}^N {\alpha_i} = 1.
\end{equation}

It is worth mentioning that this problem includes the allocation of the rates to the nodes, but is not limited to it, since rate allocation is equivalent to the specification of the determinant of the distortion matrices ${\bf{D}}_i$ according to (\ref{RDF_enhance2}), while in (\ref{MIN}) the whole matrices ${\bf{D}}_1, ..., {\bf{D}}_N$ must be chosen optimally to maximize the output SNR. This possibility is a result of having the target distortions in the matrix form.

Problem (\ref{MIN}) is equivalent to the following problem:

\begin{align}
\label{MIN1}
&\min_{{\bf{D}}_1, ..., {\bf{D}}_N}  {\text{tr}\left\{ \left[ {\bf{W}} \left( {{\bf{\Sigma}}_{\bf{v}} ({\bf{D}}_1, ..., {\bf{D}}_N)} \right) ^{-1} {\bf{W}}^T \right]^{-1} \right\}}  \nonumber  \\
&\;\;\;\; s.t. \;\;\;\;  \sum_{i=1}^N \alpha_i R({\bf{D}}_i) = R.
\end{align}

\noindent
Substituting (\ref{RDF_enhance2}) in the constraint in (\ref{MIN1}), noting that ${{\bf{\Sigma}}_{{\bf{v}}_i}} = {{\bf{\Sigma}}_{{\bf{n}}_i} } + {\bf{\Sigma }}_{{\boldsymbol{\nu}}_i}$, and using (\ref{yhat2}), we rewrite (\ref{MIN1}) as:

\begin{align}
\label{MIN2}
&\min_{{\bf{D}}_1, ..., {\bf{D}}_N} \! {\text{tr} \! \left\{ \!\! \left( \sum_{i=1}^N {\bf{W}}_i {\left[  {{\bf{\Sigma}}_{{\bf{n}}_i} } + {\left( {\bf{D}}_i^{-1} - {\bf{\Sigma}}_{{\bf{y}}_i}^{-1} \right)}^{-1}  \right]}^{-1} {\bf{W}}_i^T \! \right)^{-1} \! \right\}} \nonumber \\
&   \;\;\;\; s.t. \;\;\;\;  \prod_{i=1}^N \left| {\bf{D}}_i \right| ^{\alpha_i} = \beta,
\end{align}

\noindent
where $\beta$ is defined as:

\begin{equation}
\label{beta}
\beta = e^{-2R} \prod_{i=1}^N \left| {\bf{\Sigma}}_{{\bf{y}}_i} \right| ^{\alpha_i}.
\end{equation}

For simplicity, we assume that ${\bf{W}}_i$ in (\ref{predictablity}) are invertible. Notice that even if the relationship between ${\bf{x}}_i$ and ${\bf{x}}_d$ is not invertible, one could find an invertible matrix ${\bf{W}}_i$ which minimizes for example the mean-squared error between ${\bf{x}}_i$ and ${\bf{W}}^T_i \, {\bf{x}}_d$. In this case, (\ref{predictablity}) would be an approximation (See \cite{benesty} for more details). However, the approximation error could be added to the additive noise ${\bf{n}}_i$ in (\ref{vector_samples}). One could thus assume that ${\bf{y}}_i = {\bf{W}}^T_i \, {\bf{x}}_d + {\bf{n}}_i$ always holds with an invertible ${\bf{W}}_i$. 

We define ${\bf{Z}}_i$ and ${\bf{C}}_i$ as:

\begin{align}
\label{KKT1}
& {\bf{Z}}_i = {\bf{W}}_i {{\bf{\Sigma}}_{{\bf{n}}_i}^{-1}} \left( {{\bf{\Sigma}}_{{\bf{n}}_i}^{-1}} + {\bf{D}}_i^{-1} - {{\bf{\Sigma}}_{{\bf{y}}_i}^{-1}} \right)^{-1} {{\bf{\Sigma}}_{{\bf{n}}_i}^{-1}} {\bf{W}}^T_i, \\
\label{KKT2}
& {\bf{C}}_i = {\bf{W}}_i {{\bf{\Sigma}}_{{\bf{n}}_i}^{-1}} \left( {{\bf{\Sigma}}_{{\bf{n}}_i}^{-1}} - {{\bf{\Sigma}}_{{\bf{y}}_i}^{-1}} \right)^{-1} {{\bf{\Sigma}}_{{\bf{n}}_i}^{-1}} {\bf{W}}^T_i.
\end{align}

\noindent
The following set of equations then follow from the KKT conditions for Problem (\ref{MIN2}) (See Appendix \ref{appendix_kkt} for the derivation):

\begin{align}
\label{KKT_result1}
& \alpha_i \lambda {\bf{A}}^2 = {\bf{Z}}_i - {\bf{Z}}_i {\bf{C}}_i^{-1} {\bf{Z}}_i, \\
\label{KKT_result2}
& {\bf{A}} = \sum_{i=1}^N {{\bf{W}}_i {{\bf{\Sigma}}_{{\bf{n}}_i}^{-1}} {\bf{W}}^T_i} - \sum_{i=1}^N {{\bf{Z}}_i} \\
\label{KKT_result3}
& \prod_{i=1}^N { \left| {{\bf{\Sigma}}_{{\bf{n}}_i}^{-1}} {\bf{W}}^T_i {\bf{Z}}_i^{-1} {\bf{W}}_i {{\bf{\Sigma}}_{{\bf{n}}_i}^{-1}} - {{\bf{\Sigma}}_{{\bf{n}}_i}^{-1}} + {{\bf{\Sigma}}_{{\bf{y}}_i}^{-1}} \right| ^{-\alpha_i}} =  \beta.
\end{align}

In general, it is not easy to solve (\ref{KKT_result1})--(\ref{KKT_result3}) analytically to find the unknowns ${\bf{A}}$, $\lambda$, and ${\bf{Z}}_i; i=1,\cdots,N$. Instead, one could proceed by using numerical methods. In the rest of this subsection, we consider special cases of this problem where further analysis is possible and leads to interesting results.

\vspace{4mm}
%********************************************************************************************************************************************************
\subsubsection{High-Rate Regime}
\label{high_rate}
We assume that (\ref{KKT_result1}) could be approximated as:

\begin{equation}
\label{KKT_result1_aprx}
\alpha_i \lambda {\bf{A}}^2 \approx {\bf{Z}}_i.
\end{equation}

\noindent
From (\ref{KKT1}) and (\ref{KKT2}), this means that the distortion matrices ${\bf{D}}_i$ in (\ref{KKT1}) have small eigenvalues, or equivalently, ${\bf{D}}_i^{-1}$ has large eigenvalues. This holds, if the rates $R({\bf{D}}_i)$ are high enough. We will show later, that in fact in order for the results to hold, the rates need not be very high, and depending on the setup, even for relatively low rates the results will be valid. The high-rate assumption can also be interpreted as:

\begin{equation}
\label{other-form}
{\bf{D}}_i^{-1} + {{\bf{\Sigma}}_{{\bf{n}}_i}^{-1}} - {{\bf{\Sigma}}_{{\bf{y}}_i}^{-1}} \approx {\bf{D}}_i^{-1}.
\end{equation}

\noindent We will use this form later to find the optimal value of the parameter $\lambda$. Combining (\ref{KKT_result2}) and (\ref{KKT_result1_aprx}), using (\ref{weights}), and defining:

\begin{equation}
\label{S}
{\bf{S}} = \sum_{i=1}^N {{\bf{W}}_i {{\bf{\Sigma}}_{{\bf{n}}_i}^{-1}} {\bf{W}}^T_i}, 
\end{equation}

\noindent
yields:

\begin{equation}
\label{A_eqn}
\lambda {\bf{A}}^2 + {\bf{A}} -  {\bf{S}} \approx {\bf{0}}.
\end{equation}

\noindent
From (\ref{A_eqn}), it is easy to see that ${\bf{A}}$ and ${\bf{S}}$ share the same eigenvectors. Denote the eigenvalue decomposition of ${\bf{S}}$ by:

\begin{equation}
\label{S_eig}
{\bf{S}} = {\bf{U}}_s \text{diag}\{s_i, i=1,\cdots,n\} {\bf{U}}^T_s.
\end{equation}

\noindent
One could then write ${\bf{A}}$ as:

\begin{equation}
\label{A_eig}
{\bf{A}} \approx {\bf{U}}_s \text{diag}\{a_i, i=1,\cdots,n\} {\bf{U}}^T_s.
\end{equation}

\noindent
To find $a_i, i=1,\cdots,n$, we substitute (\ref{A_eig}) and (\ref{S_eig}) in (\ref{A_eqn}) to obtain:

\begin{equation*}
%\label{A_eqn2}
\lambda a^2_i + a_i - s_i \approx 0; \,\,\, i=1,\cdots,n,
\end{equation*}

\noindent
which gives the following solution:

\begin{equation}
\label{A_eqn3}
a_i \approx \frac{\sqrt{1+4 \lambda s_i} - 1}{2 \lambda}; \,\,\, i=1,\cdots,n.
\end{equation}

Note that the only unknown parameter in (\ref{A_eqn3}) is $\lambda$. To find $\lambda$, we use (\ref{other-form}) and (\ref{KKT1}) to rewrite (\ref{KKT_result3}) as:

\begin{equation}
\label{KKT3_aprx}
\prod_{i=1}^N { \left| {{\bf{\Sigma}}_{{\bf{n}}_i}^{-1}} {\bf{W}}^T_i {\bf{Z}}_i^{-1} {\bf{W}}_i {{\bf{\Sigma}}_{{\bf{n}}_i}^{-1}} \right| ^{-\alpha_i}} \approx \beta.
\end{equation}

\noindent
Defining $\gamma$ as:

\begin{equation}
\label{gamma}
\gamma = \beta  \prod_{i=1}^N { \left| \alpha_i {{\bf{\Sigma}}_{{\bf{n}}_i}} {\bf{W}}^{-T}_i {\bf{W}}^{-1}_i {{\bf{\Sigma}}_{{\bf{n}}_i}} \right| ^{-\alpha_i}},
\end{equation}

\noindent
substituting (\ref{KKT_result1_aprx}) in (\ref{KKT3_aprx}), and simplifying the result yields:

\begin{equation}
\label{lambda1}
\lambda^n \left| {\bf{A}}^2 \right| \approx \gamma.
\end{equation}

\noindent
Finally, substituting (\ref{A_eqn3}) in (\ref{lambda1}) leads to the following equation for $\lambda$:

\begin{equation}
\label{lambda2}
\prod_{i=1}^N { \left( \frac{\sqrt{1+4 \lambda s_i} - 1}{\sqrt{4 \lambda}} \right)^2 } \approx \gamma.
\end{equation}

To summarize the results, given the parameters ${\bf{W}}_i$, $\alpha_i$, ${{\bf{\Sigma}}_{{\bf{n}}_i}}$ for $i=1,\cdots,n$, one should take the following steps to find the optimal allocation of the distortions ${\bf{D}}_i$ to the nodes:

\begin{itemize}
\item Calculate ${\bf{S}}$ using (\ref{S}), and find ${\bf{U}}_s$ and $s_i; i=1,\cdots,n$ from (\ref{S_eig}).
\item Calculate $\gamma$ from (\ref{gamma}) and then find $\lambda$ by solving (\ref{lambda2}).
\item Use (\ref{A_eqn3}) and (\ref{A_eig}) to find ${\bf{A}}$.
\item Calculate ${\bf{Z}}_i$ using (\ref{KKT_result1_aprx}).
\item Calculate ${\bf{D}}_i$ using (\ref{KKT1}).
\end{itemize}

Note that the whole process depends on finding a solution for (\ref{lambda2}). Define $R_{min}$ as:

\begin{equation*}
%\label{Rmin}
R_{min} = \frac{1}{2} \log{ \frac{ \prod_{i=1}^N { \left| \alpha_i {{\bf{\Sigma}}_{{\bf{n}}_i}} {\bf{W}}^{-T}_i {\bf{W}}^{-1}_i {{\bf{\Sigma}}_{{\bf{n}}_i}} \right| ^{-\alpha_i} \left| {{{\bf{\Sigma}}_{{\bf{y}}_i}}} \right|^{\alpha_i} } }{\left| {\bf{S}} \right|} }.
\end{equation*}

\noindent
We then prove the following proposition:

\begin{proposition}
\label{high_rate_proposition}
If $R \geq R_{min}$, there is a unique solution to (\ref{lambda2}), otherwise (\ref{lambda2}) has no solution.
\end{proposition}

\begin{IEEEproof}
See Appendix \ref{appendix_uniqueness_highrate}.
\end{IEEEproof}

\vspace{1em}
We perform a few simulations to further study the behaviour of the output SNR as a function of the distortion matrices ${\bf{D}}_i$. We assume that there are two nodes $N = 2$, and ${\bf{W}}_1 = {\bf{W}}_2 = {\bf{I}}_n$. We consider the following structure for the covariance matrix of both signal and noise:

\begin{equation*}
{\bf{\Sigma}}_{(i,j)} = \nu \rho^{|i-j|},
\end{equation*}

\noindent
where ${\bf{\Sigma}}_{(i,j)}$ is the $(i,j)$th element of ${\bf{\Sigma}}$. For the signal ${\bf{x}}_d$, we choose $\nu = 1$ and $\rho = 0.9$. For the noise terms, we use $\nu$ and $\rho$ as control parameters. The dimension and the weighted sum-rate are kept fixed on $n=32$ and $R = 2.5n$, respectively, unless otherwise stated.

We check the optimality of the results and the gain that can be achieved. We first specify the parameters $\alpha_1,\alpha_2$, and the covariance matrix of the noise terms. We find the distortion matrices ${\bf{D}}^*_i; i=1,2$ suggested by the procedure described above and their corresponding SNR. We then compare the resulting SNR to the SNR given by other choices of the distortion matrices. Assuming that the eigenvalue decomposition of ${\bf{D}}^*_i$ is given by ${\bf{D}}^*_i = {\bf{U}}_i {\bf{\Lambda}}_i {\bf{U}}^T_i$, we use the following formula to generate other distortion matrices:

\begin{equation}
\label{random_distortion}
{\bf{D}}_i = {\bf{U}}_i \left( \beta_i {\bf{\Lambda}}_i + \eta_i {\bf{\Theta}}_i \right) {\bf{U}}^T_i,
\end{equation}

\noindent
where $\beta_i$ and $\eta_i$ are weighting parameters, and ${\bf{\Theta}}_i$ is a diagonal matrix whose diagonal elements are drawn from a uniform distribution in the interval $[0,5\iota_i]$, with $\iota_i$ being the largest element of ${\bf{\Lambda}}_i$. By choosing $\beta_i$ close to $1$ and $\eta_i$ close to $0$, the resulting distortion matrices ${\bf{D}}_i$ will be slightly perturbed versions of ${\bf{D}}^*_i$. On the other hand, by choosing $\beta_i$ close to $0$ and $\eta_i$ close to $1$, the resulting ${\bf{D}}_i$ will be random. Using (\ref{random_distortion}), we generate $L$ valid distortion pairs $({\bf{D}}_1,{\bf{D}}_2)$ using the following algorithm:

\vspace{1em}
\noindent \textbf{Initialization}: Set $j = 1$.

\noindent \textbf{Iterations}: While $j \leq L$ perform the following steps: 
\begin{enumerate}
\item Generate ${\bf{D}}_1$ using (\ref{random_distortion})
\item If the result is not a valid distortion matrix, go back to 1, otherwise Generate ${\bf{D}}_2$ using (\ref{random_distortion}).
\item Normalize ${\bf{D}}_2$ such that the constraint in (\ref{MIN}) holds. If the result is not a valid distortion matrix, go back to 2.
\item Calculate the SNR given by $({\bf{D}}_1,{\bf{D}}_2)$.
\item Increase $j$ by 1.
\end{enumerate}
\vspace{1em}

\begin{figure*}[tb]
\begin{center}
\includegraphics[scale=0.28]{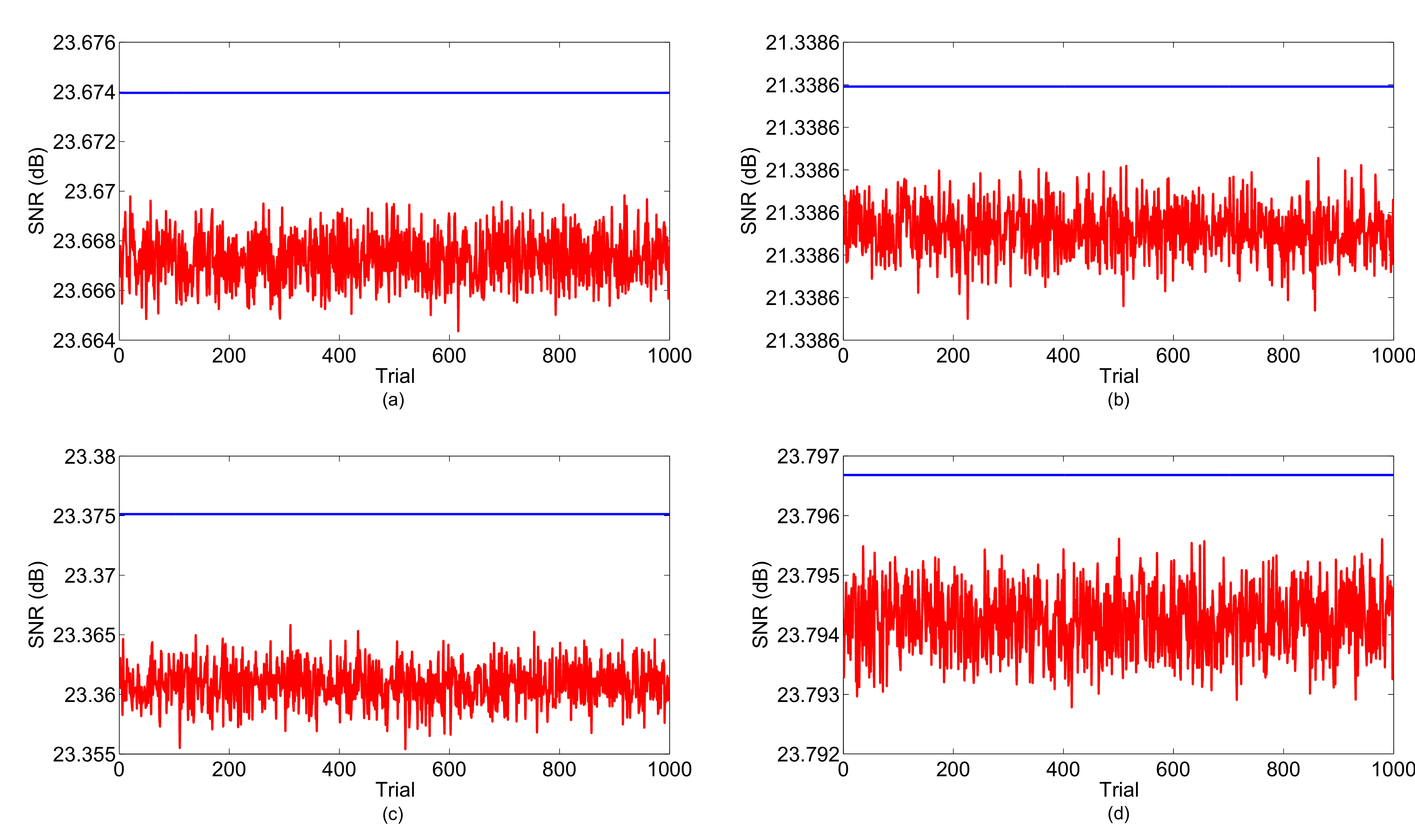}
\caption{Output SNR for 1000 trials where the distortion matrices $({\bf{D}}_1,{\bf{D}}_2)$ are slightly perturbed versions of $({\bf{D}}^*_1,{\bf{D}}^*_2)$. The blue line indicates the output SNR at $({\bf{D}}^*_1,{\bf{D}}^*_2)$. (a) with $\alpha_1=\alpha_2=0.5$, and $\rho_1=0.9,\rho_2=0.3$, $\nu_1=0.01,\nu_2=0.02$ for the covariance matrix of the noise terms, (b) $\alpha_1=\alpha_2=0.5$, and $\rho_1=\rho_2=0$, $\nu_1=0.01,\nu_2=0.02$, (c) $\alpha_1=0.7,\alpha_2=0.3$, and $\rho_1=0.9,\rho_2=0.3$, $\nu_1=0.01,\nu_2=0.02$, (d) $\alpha_1=0.3,\alpha_2=0.7$, and $\rho_1=0.9,\rho_2=0.3$, $\nu_1=0.01,\nu_2=0.02$.}
\label{local_max}
\end{center}
\vspace{-1.5em}
\end{figure*}

We set $L=10^3$ for the rest of this subsection. To study the behavior of the output SNR around $({\bf{D}}^*_1,{\bf{D}}^*_2)$, we set $\beta_1 = \beta_2 = 0.999$ and $\eta_1 = \eta_2 = 0.001$. Figure \ref{local_max} illustrates the results for different choices of the parameters. As seen from Fig. \!\ref{local_max}, the theoretical results from the high-rate approximation indeed give the local maximum for the output SNR.

\begin{figure*}[tb]
\begin{center}
\includegraphics[scale=0.28]{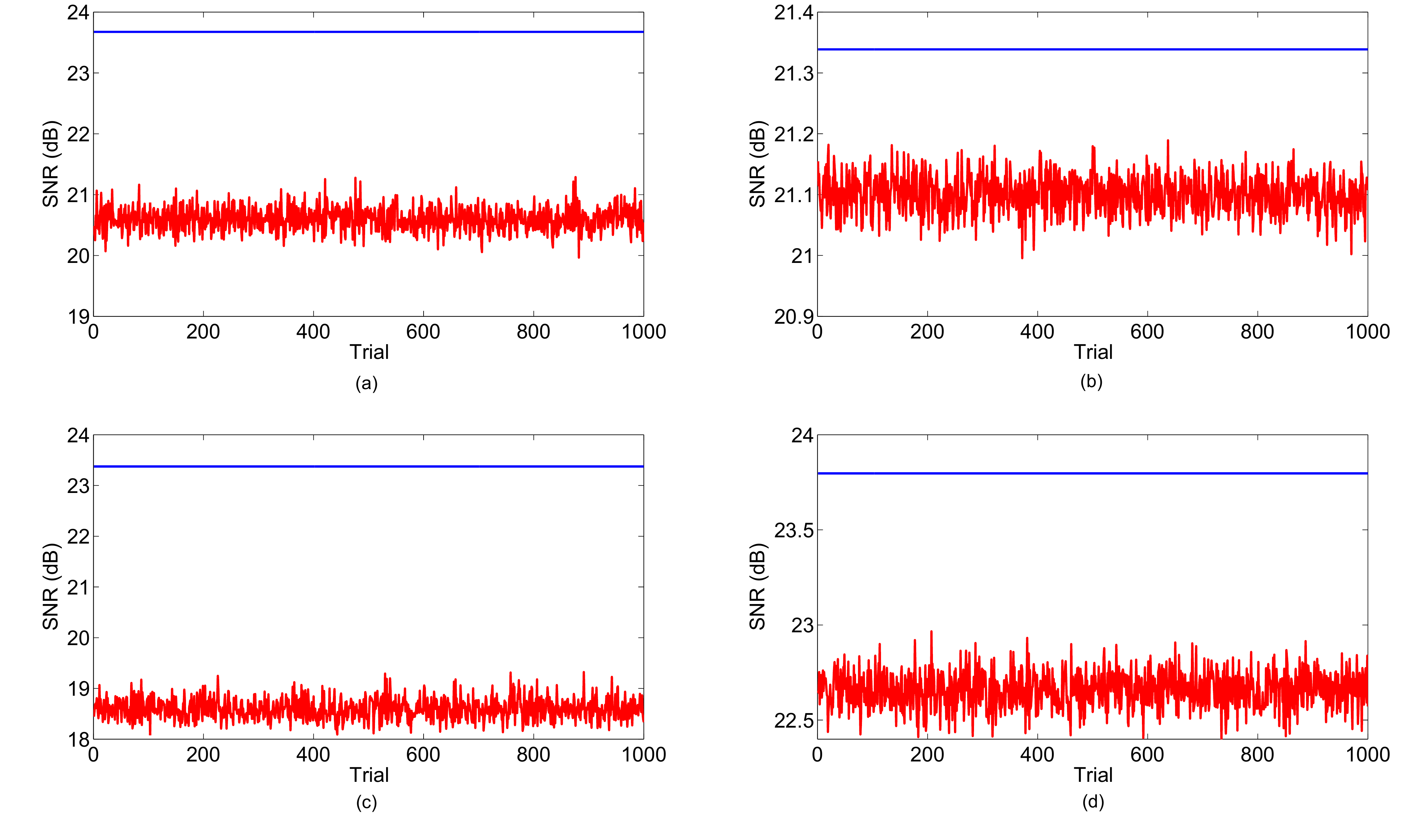}
\caption{Output SNR for 1000 trials where the distortion matrices $({\bf{D}}_1,{\bf{D}}_2)$ are chosen randomly. The blue line indicates the output SNR at $({\bf{D}}^*_1,{\bf{D}}^*_2)$. (a) with $\alpha_1=\alpha_2=0.5$, and $\rho_1=0.9,\rho_2=0.3$, $\nu_1=0.01,\nu_2=0.02$ for the covariance matrix of the noise terms, (b) $\alpha_1=\alpha_2=0.5$, and $\rho_1=\rho_2=0$, $\nu_1=0.01,\nu_2=0.02$, (c) $\alpha_1=0.7,\alpha_2=0.3$, and $\rho_1=0.9,\rho_2=0.3$, $\nu_1=0.01,\nu_2=0.02$, (d) $\alpha_1=0.3,\alpha_2=0.7$, and $\rho_1=0.9,\rho_2=0.3$, $\nu_1=0.01,\nu_2=0.02$.}
\label{global_max}
\end{center}
\vspace{-1.5em}
\end{figure*}

To verify that $({\bf{D}}^*_1,{\bf{D}}^*_2)$ is the global maximizer of the output SNR and to see the gap between the maximum SNR and the one resulting from a random allocation of the distortion matrices, we set $\beta_1 = \beta_2 = 0$ and $\eta_1 = \eta_2 = 1$. The results are illustrated in Fig. \!\ref{global_max}. As seen from the plots, the gap between the optimal SNR and the one resulting from a random allocation could in some cases be as large as 5 dB.

\begin{figure}[tb]
\begin{center}
\includegraphics[scale=0.2]{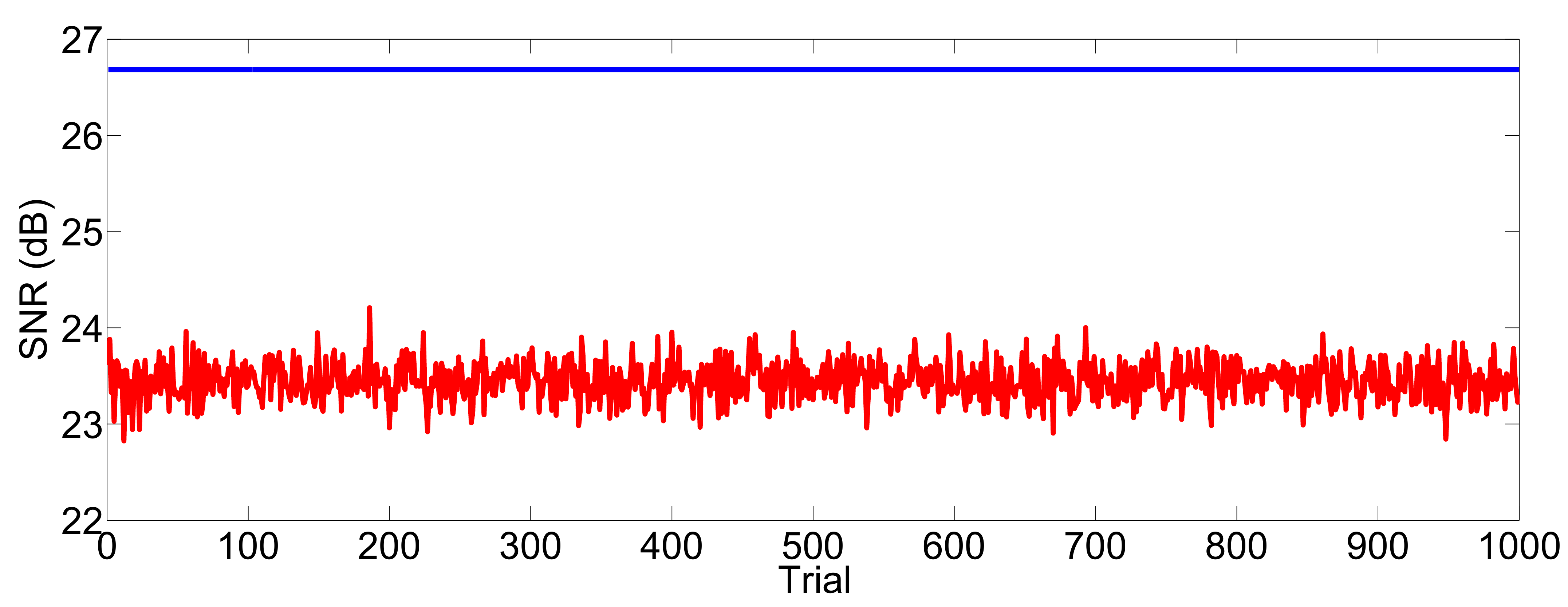}
\caption{Output SNR at 1000 trials for a simulation where there are 4 sensors and the distortion matrices $({\bf{D}}_1,{\bf{D}}_2,{\bf{D}}_3,{\bf{D}}_4)$ are chosen randomly. The blue line indicates the output SNR at $({\bf{D}}^*_1,{\bf{D}}^*_2,{\bf{D}}^*_3,{\bf{D}}^*_4)$, which gives the optimal solution. The choice of parameters is similar to the simulation in Fig. \!\ref{global_max} (a).}
\label{4nodes}
\end{center}
\end{figure}

To show the scalability in sense of the number of sensors, we repeat the simulation in Fig. \!\ref{global_max} (a) with 4 sensors. Two sensors have similar parameters with sensor 1 in Fig. \!\ref{global_max} (a), and the other two sensors have similar parameters with sensor 2 in the same figure. The weights are set as $\alpha_1=\alpha_2=\alpha_3=\alpha_4=0.25$. The result is shown in Fig. \ref{4nodes}, which shows that for 4 sensors, as expected, the output SNR is higher compared to 2 sensors (Fig. \ref{global_max} (a)) due to the higher number of the available noisy signals. Moreover, comparison of the figures shows that the gap between the optimal performance and the one resulting from the random allocation of the distortion matrices is similar for the 4 sensors and 2 sensors scenarios.

\begin{figure}[tb]
\begin{center}
\includegraphics[scale=0.18]{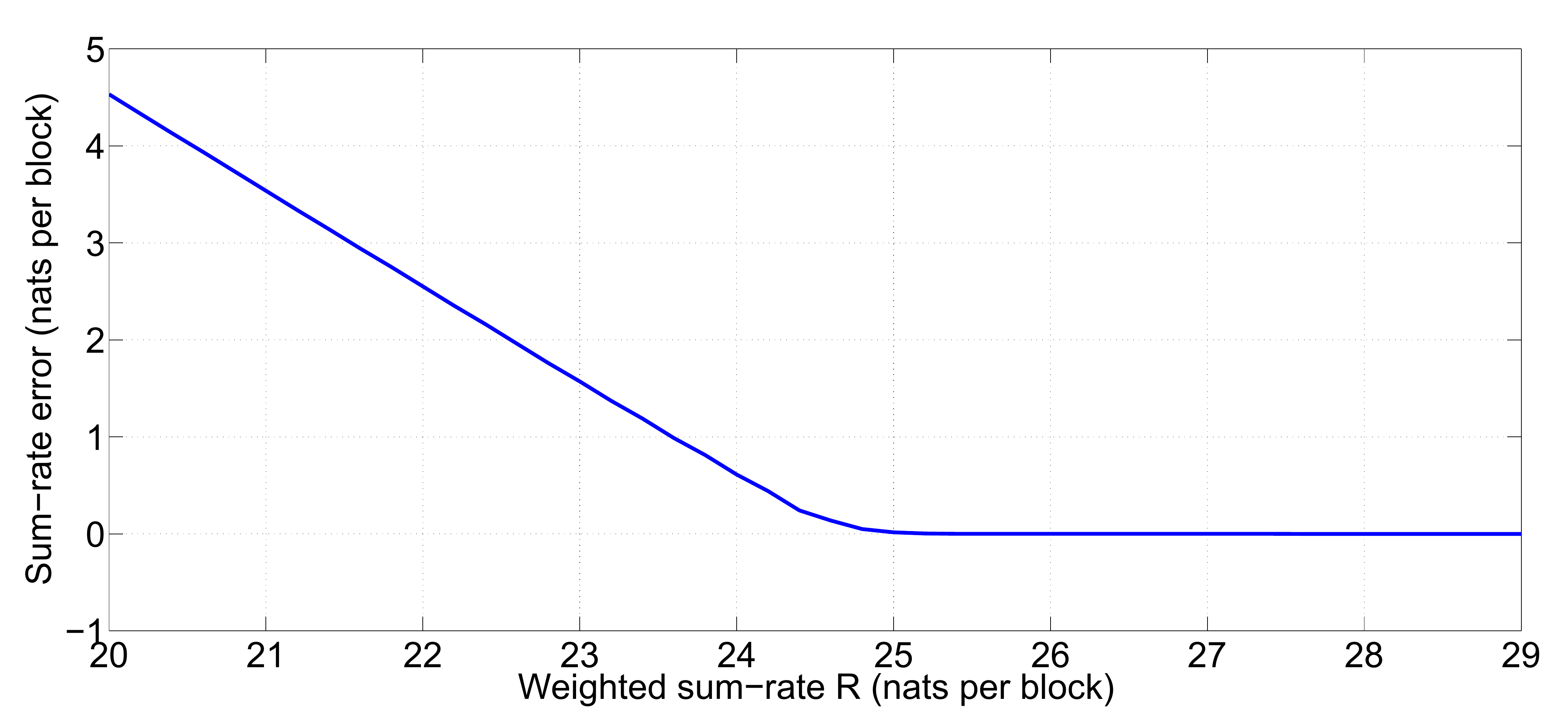}
\caption{The difference between the desired weighted sum-rate $R$ and the resulting weighted sum-rate due to the high-rate approximation as a function of $R$}
\label{accuracy}
\end{center}
\vspace{-1.5em}
\end{figure}

Next we study the validity of the high-rate assumption. Notice that even with low rates, the procedure described to find the optimal distortion allocation might give a solution. The problem, however is that if the rates are not high enough, the approximation in (\ref{KKT_result1_aprx}) will not hold, and the resulting distortion matrices might violate the sum-rate constraint in (\ref{MIN}). The approximation error is negligible if $\alpha_1 R({\bf{D}}^*_1) + \alpha_2 R({\bf{D}}^*_2) \approx R$. The difference between $\alpha_1 R({\bf{D}}^*_1) + \alpha_2 R({\bf{D}}^*_2)$ and $R$ as a function of $R$ is plotted in Fig. \!\ref{accuracy} for $\alpha_1=\alpha_2=0.5$, and $\rho_1 = \rho_2 = 0.8$ and $\nu_1 = 0.1,\nu_2=0.2$ for the covariance matrices of the noise terms. Note that for a weighted sum-rate $R$ that is as low as $25$ nats per block (equivalent to $0.78$ nats per sample or $1.13$ bits per sample) the high-rate assumption still holds with a negligible error.

\vspace{4mm}
%********************************************************************************************************************************************************
\subsubsection{Scalar Sources}
\label{scalar}
We assume that $n = 1$, which means that all the sources are scalar. We further assume that there are only two nodes. While the high-rate case studied in the previous subsection is of interest for practical applications, the scalar case with two nodes may be of less practical relevance. However, it leads to interesting analytical results that makes it worth studying.

We denote the scalar version of ${\bf{D}}_i$ and ${\bf{W}}_i$ by $D_i$ and $w_i$, respectively. The variance of random variables is denoted by $\Sigma$ followed by a subscript. We assume that we are interested in the sum-rate (so $\alpha_1 = \alpha_2 = 0.5$) and $w_1 = w_2$ to further simplify the problem for analytical derivations. Applying these assumptions to (\ref{KKT1})--(\ref{KKT_result3}), and defining $\Sigma_1$ and $\Sigma_2$ as:

\begin{equation*}
%\label{sig}
\Sigma_i = \Sigma_{n_i} - \frac{{\Sigma_{n_i}}^2}{\Sigma_{y_i}}, \;\;\;\; i = 1,2,
\end{equation*}

\noindent
one can show that the target distortions $D^*_1$ and $D^*_2$ are given by:

\begin{align}
\label{d1}
& D^*_1 = \beta \frac{\Sigma_{n_1}}{\Sigma_{n_2}} \left( \frac{\Sigma_{n_1}\Sigma_{n_2} - \Sigma_2 \beta}{\Sigma_{n_1}\Sigma_{n_2} - \Sigma_1 \beta} \right),\\
\label{d2}
& D^*_2 = \beta \frac{\Sigma_{n_2}}{\Sigma_{n_1}} \left( \frac{\Sigma_{n_1}\Sigma_{n_2} - \Sigma_1 \beta}{\Sigma_{n_1}\Sigma_{n_2} - \Sigma_2 \beta} \right).
\end{align}

\noindent
Based on this, and defining the following parameters:

\begin{align}
\label{r1}
& R_{max} = \frac{1}{4} \log \frac{\left[\max(\Sigma_1,\Sigma_2)\right]^2}{(\Sigma_{n_1} - \Sigma_1)(\Sigma_{n_2} - \Sigma_2)},\\
\label{r2}
& R_{min} = \frac{1}{4} \log \frac{\left[\min(\Sigma_1,\Sigma_2)\right]^2}{(\Sigma_{n_1} - \Sigma_1)(\Sigma_{n_2} - \Sigma_2)},
\end{align}

\noindent
we prove the following proposition.

\begin{proposition}
\label{prop}
The stationary point $(D^*_1,D^*_2)$ given by (\ref{d1})-(\ref{d2}) is the unique maximizer of the output SNR, if $R > R_{max}$, and is the unique minimizer, if $R < R_{min}$.
\end{proposition}

\begin{IEEEproof}
See Appendix \ref{appendix_scalar}
\end{IEEEproof}

\begin{figure}[tb]
\begin{center}
\includegraphics[scale=0.21]{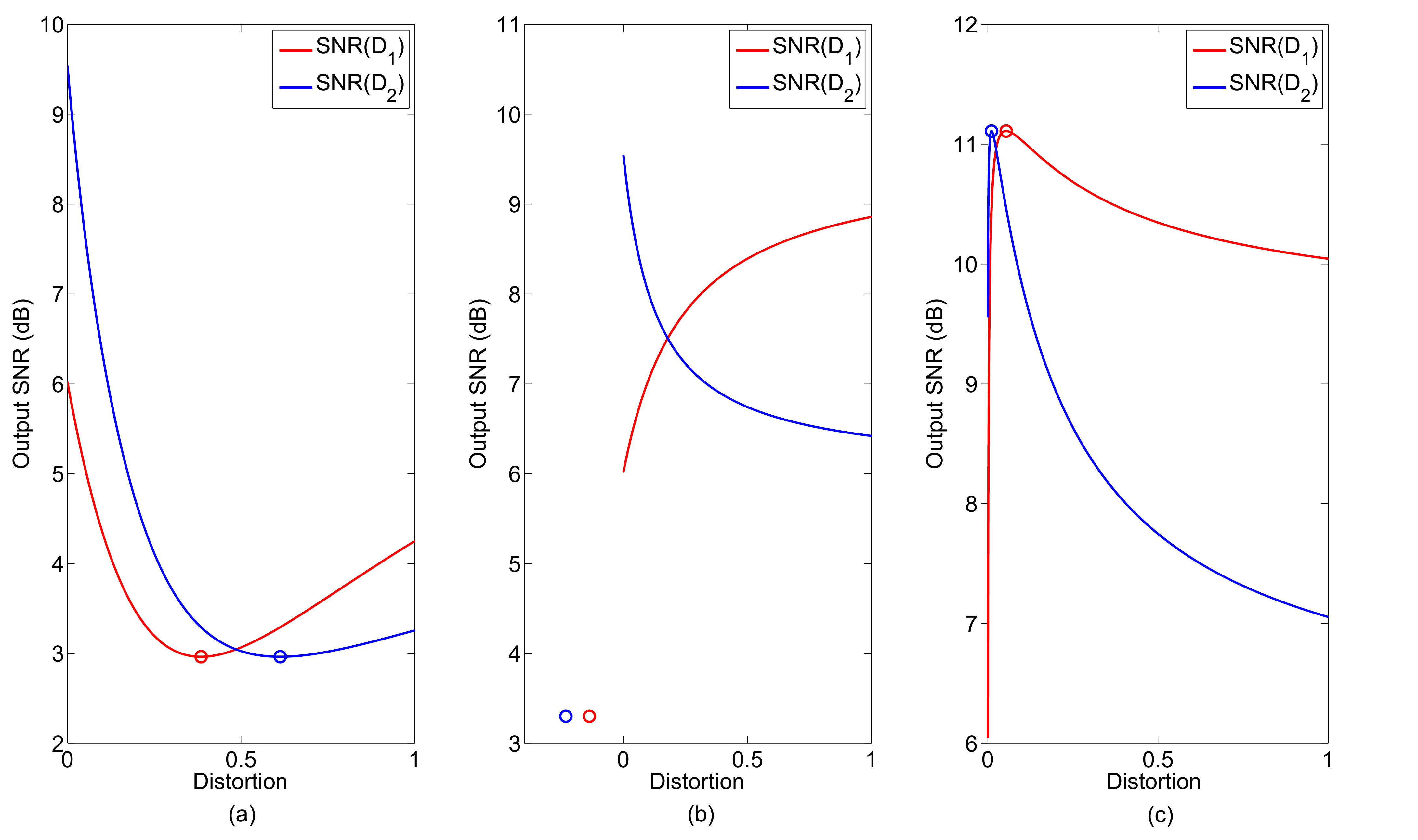}
\caption{An example of the output SNR as a function of $D_1$ and $D_2$ for the scalar case with two nodes for three different values of the network sum-rate, (a) $R<R_{min}$, (b) $R_{min}<R<R_{max}$, and (c) $R>R_{max}$. The circles show the stationary points suggested by the KKT conditions.}
\label{plots}
\end{center}
\vspace{-1.5em}
\end{figure}

\vspace{1em}
To observe the behaviour of the output SNR as a function of the distortions, consider the following example. Let $w_1 = w_2 = 1$, $\Sigma_{x_d} = 1$, $\Sigma_{y_1} = 1.2$, $\Sigma_{n_1} = 0.2$, $\Sigma_{y_2} = 1.1$, and $\Sigma_{n_2} = 0.1$. From (\ref{r1}) and (\ref{r2}) we have $R_{max} = 1.13$ and $R_{min} = 0.83$. We make use of the weighted sum-rate constraint $\frac{1}{2} R(D_1) + \frac{1}{2} R(D_2) = R$ to write $\text{SNR}(D_1,D_2)$ as a function of $D_1$ or $D_2$ only (denoted by $\text{SNR}(D_1)$ and $\text{SNR}(D_2)$, respectively). We plot $\text{SNR}(D_1)$ and $\text{SNR}(D_2)$ for three different values $0.5, 1, 2$ of the weighted sum-rate $R$. The result is shown in Fig. \!\ref{plots}. As seen in Fig. \!\ref{plots} \!(a), for $R=0.5<R_{min}$ the output SNR is minimized at $(D_1 = D^*_1,D_2 = D^*_2)$, which is not desired. The maximum SNR should then be on the boundary, which means that the whole rate should be given to one of the nodes. It can be seen in the figure that the highest SNR is achieved for the smallest $D_2$, suggesting that the whole rate should be allocated to node 2. This is in agreement with intuition, since node 2 is less noisy than node 1. For $R_{min}<R<R_{max}$, one could see in Fig. \!\ref{plots} \!(b) that $(D_1 = D^*_1,D_2 = D^*_2)$ is not a feasible point. Similar to the previous case, the optimal solution should again be on the boundary. Finally, for $R=2>R_{max}$, $(D_1 = D^*_1,D_2 = D^*_2)$ maximizes the output SNR.

Note that the above example suggests that the distortion allocation has a water-filling form. The critical rate $R_{max}$ acts as the water level. If $R$ is above this level, the rate is split between the nodes based on their noise levels. If $R$ is below the water level, the noisier node is omitted, and the whole rate is given to the other node.

%********************************************************************************************************************************************************
%********************************************************************************************************************************************************
%********************************************************************************************************************************************************
%********************************************************************************************************************************************************

\section{Conclusions and Discussion}
\label{conclusion}

We considered a source coding problem in a networked setup under covariance matrix distortion constraints. We modelled the problem as a vector Gaussian remote Wyner-Ziv problem and solved the problem by deriving an explicit formula for the rate-distortion function and designing coding schemes that asymptotically achieve the rate-distortion function. We then studied some applications of the results. In particular, we showed that the rate-distortion function for the equivalent Wyner-Ziv problem with mean-squared error distortions and the rate-information function modelling a relay network source coding problem are special cases of our results. Finally, we considered a centralized sensor network with a weighted sum-rate constraint where each node transmits its observation with a certain rate and distortion, and the received data is fused at the center to maximize the output SNR without enforcing linear distortions. For this problem we bridged between noise reduction and source coding by showing that the distortion matrices and the rates at the individual nodes could be designed to maximize the output SNR at the center. We considered special cases such as the high-rate case or the case of scalar sources in order to obtain analytical results. Further work could be other possible special cases of the noise reduction problem where analytical results can be obtained.

%********************************************************************************************************************************************************
%********************************************************************************************************************************************************
%********************************************************************************************************************************************************
%********************************************************************************************************************************************************

\appendices

%********************************************************************************************************************************************************
\section{}
\label{appendix2}

From the fact that ${\bf{n}}$ is uncorrelated with ${\bf{y}}$ and ${\bf{z}}$ and:

\begin{equation}
\label{xyz_vect}
\bf{x}=  \left( \!\! \begin{array}{cc}
{\bf{A}} & {\bf{B}} \\
\end{array} \!\! \right)
\left( \!\! \begin{array}{c}
{\bf{y}} \\
{\bf{z}} \\
\end{array} \!\! \right)
+ {\bf{n}},
\end{equation}

\noindent
it follows that:

\begin{align*}
%\label{xyz_result}
& \left( \!\! \begin{array}{cc}
{\bf{A}} & {\bf{B}} \\
\end{array} \!\! \right) = 
\left( \!\! \begin{array}{cc}
{{{\bf{\Sigma }}_{\bf{xy}}}} & {{{\bf{\Sigma }}_{\bf{xz}}}} \\
\end{array} \!\! \right)
\left( \!\! \begin{array}{ll}
{{{{\bf{\Sigma }}_{\bf{y}}}}} & {{{{\bf{\Sigma }}_{\bf{yz}}}}} \\
{{{{\bf{\Sigma }}_{\bf{zy}}}}} & {{{{\bf{\Sigma }}_{\bf{z}}}}} \\
\end{array} \!\! \right)^{-1} \\
& =\!\! \left( \!\!\! \begin{array}{cc}
{{{\bf{\Sigma }}_{\bf{xy}}} {\bf{\Delta}}_1^{\!\!-1} \!\!-\!\! {{\bf{\Sigma }}_{\bf{xz}}} {\bf{\Delta}}_2^{\!\!-1} {{\bf{\Sigma }}_{\bf{zy}}} {\bf{\Sigma}}_{\bf{y}}^{-1}} & \!\! {{{\bf{\Sigma }}_{\bf{xz}}} {\bf{\Delta}}_2^{\!\!-1} \!\!-\!\! {{\bf{\Sigma }}_{\bf{xy}}} {\bf{\Delta}}_1^{\!\!-1} {{\bf{\Sigma }}_{\bf{yz}}} {\bf{\Sigma}}_{\bf{z}}^{-1}} \\
\end{array} \!\!\!\! \right)\!,
\end{align*}

\noindent
where ${\bf{\Delta}}_1$ and ${\bf{\Delta}}_2$ are defined as:

\begin{align*}
{\bf{\Delta}}_1 = {{\bf{\Sigma }}_{\bf{y}}} - {{\bf{\Sigma }}_{\bf{yz}}} {\bf{\Sigma}}_{\bf{z}}^{-1} {{\bf{\Sigma }}_{\bf{zy}}}, \,\,\,\, {\bf{\Delta}}_2 = {{\bf{\Sigma }}_{\bf{z}}} - {{\bf{\Sigma }}_{\bf{zy}}} {\bf{\Sigma}}_{\bf{y}}^{-1} {{\bf{\Sigma }}_{\bf{yz}}}.
\end{align*}

%********************************************************************************************************************************************************
\section{Proof of Lemma \ref{lem1}}
\label{appendix_lowerbound}

We start from the following chain of inequalities:

\begin{align}
R(\bf{D} \! ) & \! = \! \min_{\bf{u}} I \! \left(\bf{y};\bf{u}|\bf{z}\right) \text{  s.t.  } {{\bf{\Sigma }}_{{\bf{x}}|{\bf{u}}{{\bf{z}}}}} \preceq \bf{D},{\bf{u}} \! \leftrightarrow \! {\bf{y}} \! \leftrightarrow \! ({\bf{x}},{\bf{z}})  \nonumber \\
\label{lb2}
& \! \geq \! \min_{\bf{u}} I \! \left(\bf{y};\bf{u}|\bf{z}\right) \text{  s.t.  } {{\bf{\Sigma }}_{{\bf{x}}|{\bf{u}}{{\bf{z}}}}} \preceq \bf{D},{\bf{u}} \! \leftrightarrow \! ({\bf{y}},{\bf{z}}) \! \leftrightarrow \! {\bf{x}}  \\
\label{lb3}
& \! = \! \min_{\bf{u}} I \! \left(\bf{y};\bf{u}|\bf{z}\right) \text{  s.t.  } {{\bf{\Sigma }}_{{\bf{x}}|{\bf{u}}{{\bf{z}}}}} \preceq \bf{D},{\bf{u}} \! \leftrightarrow \! ({\bf{y}}',{\bf{z}}) \! \leftrightarrow \! {\bf{x}}  \\
\label{lb4}
& \! \geq \! \min_{\bf{u}} I \! \left({\bf{y}}';\bf{u}|\bf{z}\right) \!\text{  s.t.  } {{\bf{\Sigma }}_{{\bf{x}}|{\bf{u}}{{\bf{z}}}}} \!\preceq \!\bf{D},{\bf{u}} \! \leftrightarrow \! ({\bf{y}}',{\bf{z}}) \! \leftrightarrow \! {\bf{x}}  \\
& \! = \! \min_{\bf{u}} h \! \left({\bf{y}}'|\bf{z}\right) \! - \! h \! \left({\bf{y}}'|\bf{u},\bf{z}\right) \text{  s.t.  } \left\{ \!\!\! \begin{array}{ll} 
{{\bf{\Sigma }}_{{\bf{x}}|{\bf{u}}{{\bf{z}}}}} \preceq \bf{D}, \\
{\bf{u}} \! \leftrightarrow \! ({\bf{y}}',{\bf{z}}) \! \leftrightarrow \! {\bf{x}} 
\end{array} \right. \nonumber \\
\label{lb5}
& \! \geq \! \min_{{{\bf{\Sigma }}_{{\bf{y}}'|{{\bf{uz}}}}}} \! \frac{1}{2} \! \log \! \frac{\left| {{\bf{\Sigma }}_{{\bf{y}}'|{{\bf{z}}}}}  \right|}{\left| {{\bf{\Sigma }}_{{\bf{y}}'|{{\bf{uz}}}}} \right|}   \!\text{  s.t. } \!\!\left\{ \!\!\! \begin{array}{ll} 
{{\bf{\Sigma }}_{{\bf{y}}'|{{\bf{uz}}}}} \preceq \! {{\bf{\Sigma }}_{{\bf{y}}'|{{\bf{z}}}}}\\
{{\bf{\Sigma }}_{{\bf{x}}|{\bf{u}}{{\bf{z}}}}} \preceq \! \bf{D} \\
{\bf{u}} \! \leftrightarrow \! ({\bf{y}}',{\bf{z}}) \! \leftrightarrow \! {\bf{x}}
\end{array} \right. \\
& \! \geq \! \min_{{{\bf{\Sigma }}_{{\bf{y}}'|{{\bf{uz}}}}}} \! \frac{1}{2} \! \log \! \frac{\left| {{\bf{\Sigma }}_{{\bf{y}}'|{{\bf{z}}}}}  \right|}{\left| {{\bf{\Sigma }}_{{\bf{y}}'|{{\bf{uz}}}}} \right|}  \text{  s.t.  } {{\bf{\Sigma }}_{{\bf{y}}'|{{\bf{uz}}}}} \preceq \! {{\bf{\Sigma }}_{{\bf{y}}'|{{\bf{z}}}}}, {{\bf{\Sigma }}_{{\bf{x}}|{\bf{u}}{{\bf{z}}}}} \preceq \! \bf{D} \nonumber \\
\label{lb6}
& \! = \!\! \min_{{{\bf{\Sigma }}_{{\bf{y}}'|{{\bf{uz}}}}}} \!\! \frac{1}{2} \! \log \! \frac{\!\left| {{\bf{\Sigma }}_{{\bf{y}}'|{{\bf{z}}}}}  \right|}{\!\left| {{\bf{\Sigma }}_{{\bf{y}}'|{{\bf{uz}}}}} \right|}  \!\text{  s.t. } \!\!\left\{ \!\!\! \begin{array}{ll} 
\! {{\bf{\Sigma }}_{{\bf{y}}'\!|{{\bf{uz}}}}} \!\preceq \!{{\bf{\Sigma }}_{\bf{x|z}}} \!\!-\! {{\bf{\Sigma }}_{\bf{x|yz}}}\\
\!{{\bf{\Sigma }}_{{\bf{y}}'|{{\bf{uz}}}}} \!\preceq \!{\bf{D}} \!-\! {{\bf{\Sigma }}_{\bf{x|yz}}} 
\end{array} \right. \\
\label{lb7}
& = \frac{1}{2} \! \log \! \frac{\left| {{\bf{\Sigma }}_{{\bf{y}}'|{{\bf{z}}}}}  \right|}{a^*}, %\tilde{R} ({\bf{D}})
\end{align}

\noindent
where (\ref{lb2}) is because the knowledge of the side information ${\bf{z}}$ at the encoder cannot increase the rate, (\ref{lb3}) is because from (\ref{xyz}), when ${\bf{z}}$ is given, ${\bf{y}}'$ is a sufficient statistic of ${\bf{y}}$ for the estimation of ${\bf{x}}$, (\ref{lb4}) follows from (\ref{yprime}) and the data processing inequality, (\ref{lb5}) is because choosing ${{\bf{y}}'|{{\bf{uz}}}}$ to be Gaussian maximizes the differential entropy, and we have added the constraint ${{\bf{\Sigma }}_{{\bf{y}}'|{{\bf{uz}}}}} \preceq \! {{\bf{\Sigma }}_{{\bf{y}}'|{{\bf{z}}}}}$, since any valid ${{\bf{\Sigma }}_{{\bf{y}}'|{{\bf{uz}}}}}$ must satisfy this condition, (\ref{lb6}) follows from (\ref{concov1})--(\ref{concov2}), and $a^*$ is defined as:
\vspace{-3mm}

\begin{align*}
%\label{}
a^* = \max_{{{\bf{\Sigma }}_{{\bf{y}}'|{{\bf{uz}}}}}} \! {\left| {{\bf{\Sigma }}_{{\bf{y}}'|{{\bf{uz}}}}} \right|}  \text{  s.t.  } \left\{ \!\!\! \begin{array}{ll} 
{{\bf{\Sigma }}_{{\bf{y}}'|{{\bf{uz}}}}} \preceq {\bf{D}} - {{\bf{\Sigma }}_{\bf{x|yz}}}, \\
{{\bf{\Sigma }}_{{\bf{y}}'|{{\bf{uz}}}}} \preceq {{\bf{\Sigma }}_{\bf{x|z}}} - {{\bf{\Sigma }}_{\bf{x|yz}}}.
\end{array} \right.
\end{align*}

\noindent
From Lemma \ref{lem_properties2} it follows that $a^* = {\left| \min ({\bf{D}} - {{\bf{\Sigma }}_{\bf{x|yz}}} , {{\bf{\Sigma }}_{\bf{x|z}}} - {{\bf{\Sigma }}_{\bf{x|yz}}}) \right|}$. Substituting this and (\ref{concov1}) in (\ref{lb7}) yields:

\begin{equation*}
%\label{lb1}
R({\bf{D}}) \geq \frac{1}{2} \log \frac{\left| {{\bf{\Sigma }}_{\bf{x|z}}} - {{\bf{\Sigma }}_{\bf{x|yz}}} \right|}{\left| \min ({\bf{D}} - {{\bf{\Sigma }}_{\bf{x|yz}}} , {{\bf{\Sigma }}_{\bf{x|z}}} - {{\bf{\Sigma }}_{\bf{x|yz}}}) \right|}.
\end{equation*}

%********************************************************************************************************************************************************
\section{Proof of Lemma \ref{lem3}}
\label{appendix_upperbound}

Based on the results in \cite{SW,WZ}, it is enough to show that for ${\bf{u}}^*$ defined in (\ref{scheme}), we have ${I\left( {{\bf{y}};{\bf{u}}^*|{\bf{z}}} \right)} = \tilde{R} ({\bf{D}})$, and the covariance matrix ${{\bf{\Sigma }}_{\bf{x|u^*z}}}$ of the reconstruction error satisfies ${{\bf{\Sigma }}_{\bf{x|u^*z}}} \preceq {\bf{D}}$. We start with the following chain of equalities:

\begin{align}
{I\left( {{\bf{y}};{\bf{u}}^*|{\bf{z}}} \right)} & = h({{\bf{u}}^*}|{\bf{z}}) - h({{\bf{u}}^*}|{\bf{y}},{\bf{z}}) \nonumber \\
& = \frac{1}{2}\log \left( {\frac{{\left| {{\bf{\Sigma }}_{\bf{u^*|z}}} \right|}}{{\left| {{\bf{\Sigma }}_{\bf{u^*|yz}}} \right|}}} \right) \nonumber \\
\label{ub3}
& = \frac{1}{2}\log \left( {\frac{{\left|  {{\bf{U}}} {{\bf{\Sigma }}_{{\bf{y}}'|{{\bf{z}}}}} {{\bf{U}}^T} + {{{\bf{\Sigma }}_{\bf{\nu}}}} \right|}}{{\left| {{{\bf{\Sigma }}_{\bf{\nu}}}} \right|}}} \right)  \\
\label{ub4}
& = \frac{1}{2}\log \left( {\frac{{\left| {{\bf{\Lambda }} + {{\bf{\Sigma }}_{\bf{\nu}}}} \right|}}{{\left| {{{\bf{\Sigma }}_{\bf{\nu}}}} \right|}}} \right).
\end{align}

\noindent
where (\ref{ub3}) follows from (\ref{scheme}) and (\ref{yprime}), and (\ref{ub4}) follows from (\ref{concov1}). Rewriting ${\bf{\Lambda }}$ as:

\begin{align}
\label{ub6}
{\bf{\Lambda }} =  {{\bf{U}}} {( {{\bf{\Sigma }}_{\bf{x|z}}} - {{\bf{\Sigma }}_{\bf{x|yz}}} )} {{\bf{U}}^T} = {{\bf{U}}} {{\bf{V}}^{-1}} {\bf{\Lambda }} {{\bf{V}}^{-T}} {{\bf{U}}^T},
\end{align}

\noindent
and substituting (\ref{ub6}) and (\ref{cov_noise}) in (\ref{ub4}) and simplifying the result, we get:

\begin{align*}
& {I\left( {{\bf{y}};{\bf{u}}^*|{\bf{z}}} \right)} \nonumber \\
%\label{ub7}
& = \frac{1}{2}\log \left( {\frac{{\left| {{\bf{V}}^{-1}} {\bf{\Lambda }} {{\bf{V}}^{-T}} \right|}}{{\left| {{\bf{V}}^{-1}} {{\rm{diag}}\left\{ {{ \min \left({\lambda _i},{\lambda' _i} \right)} } , i=1,\dotsc,n_x \right\}} {{\bf{V}}^{-T}} \right|}}} \right)  \\
%\label{ub8}
& = \frac{1}{2} \log \frac{\left| {{\bf{\Sigma }}_{\bf{x|z}}} - {{\bf{\Sigma }}_{\bf{x|yz}}} \right|}{\left| \min ({\bf{D}} - {{\bf{\Sigma }}_{\bf{x|yz}}} , {{\bf{\Sigma }}_{\bf{x|z}}} - {{\bf{\Sigma }}_{\bf{x|yz}}}) \right|} = \tilde{R} ({\bf{D}}).
\end{align*}

Next, we derive the covariance matrix of the reconstruction error. Similar to (\ref{xyz}), we can write $\bf{x}$ as:

\begin{equation}
\label{xuz}
{\bf{x}} = {\bf{Cu^*+Gz}}+{\bf{n}}_1,
\end{equation}

\noindent
where ${\bf{n}}_1$ is independent of ${\bf{u}}^*$ and ${\bf{z}}$. Form (\ref{xuz}) it follows that ${{\bf{\Sigma }}_{\bf{xu^*|z}}} = {\bf{C}}{{\bf{\Sigma }}_{\bf{u^*|z}}}$, or:

\begin{equation}
\label{weird1}
{\bf{C}} = {{\bf{\Sigma }}_{\bf{xu^*|z}}}{{\bf{\Sigma }}^{-1}_{\bf{u^*|z}}}.
\end{equation}

\noindent
From (\ref{scheme}), (\ref{yprime}), and (\ref{concov1}) we have:

\begin{equation}
\label{weird2}
{{\bf{\Sigma }}_{\bf{u^*|z}}} = {\bf{\Lambda }} + {{\bf{\Sigma }}_\nu}.
\end{equation}

\noindent
Also note that:

\begin{eqnarray}
\label{weird3}
{{\bf{\Sigma }}_{\bf{xu^*|z}}} &&= {{\bf{\Sigma }}_{\bf{xy|z}}}{{\bf{A}}^T}{{\bf{U}}^T}\\
\label{weird4}
 &&= {\bf{A}}{{\bf{\Sigma }}_{\bf{y|z}}}{{\bf{A}}^T}{{\bf{U}}^T}\\
\label{weird5}
 &&= {{\bf{\Sigma }}_{\bf{y'|z}}} {{\bf{U}}^T} \\
\label{weird6}
 &&= \left( {{{\bf{\Sigma }}_{\bf{x|z}}} - {{\bf{\Sigma }}_{\bf{x|yz}}}} \right){{\bf{U}}^T},
\end{eqnarray}

\noindent
where (\ref{weird3}), (\ref{weird4}), (\ref{weird5}) and (\ref{weird6}) follow from (\ref{scheme}), (\ref{xyz}), (\ref{yprime}) and (\ref{concov1}), respectively. The covariance matrix of the reconstruction error can then be written as:

\begin{align}
\label{err1}
& {{\bf{\Sigma }}_{\bf{x|u^*z}}} = {{\bf{\Sigma }}_{{{\bf{n}}_1}}}\\
\label{err2}
 &= {{\bf{\Sigma }}_{\bf{x|z}}} - {\bf{C}}{{\bf{\Sigma }}_{\bf{u^*|z}}}{{\bf{C}}^T}\\
\label{err3}
 &= {{\bf{\Sigma }}_{\bf{x|z}}} - {{\bf{\Sigma }}_{\bf{xu^*|z}}}{{\bf{\Sigma }}^{ - 1}_{\bf{u^*|z}}}{{\bf{\Sigma }}^T_{\bf{xu^*|z}}}\\
\label{err4}
&= {{\bf{\Sigma }}_{\bf{x|z}}}\! -\! \left( {{{\bf{\Sigma }}_{\bf{x|z}}} \!-\! {{\bf{\Sigma }}_{\bf{x|yz}}}} \right)\!{{\bf{U}}^T}{\left( {{\bf{\Lambda }} \!+\! {{\bf{\Sigma }}_{\bf{\nu}}}} \right)^{ - 1}}{\bf{U}}\!\left( {{{\bf{\Sigma }}_{\bf{x|z}}} \!-\! {{\bf{\Sigma }}_{\bf{x|yz}}}} \right) \\
 &= {{\bf{\Sigma }}_{\bf{x|yz}}} + \left( {{{\bf{\Sigma }}_{\bf{x|z}}} - {{\bf{\Sigma }}_{\bf{x|yz}}}} \right) \nonumber \\
 & \;\;\;\;\; - \left( {{{\bf{\Sigma }}_{\bf{x|z}}} - {{\bf{\Sigma }}_{\bf{x|yz}}}} \right)\!{{\bf{U}}^T}{\left( {{\bf{\Lambda }} + {{\bf{\Sigma }}_{\bf{\nu}}}} \right)^{ - 1}}{\bf{U}}\left( {{{\bf{\Sigma }}_{\bf{x|z}}} - {{\bf{\Sigma }}_{\bf{x|yz}}}} \right) \nonumber \\
 &= {{\bf{\Sigma }}_{\bf{x|yz}}} + {{\bf{V}}^{-1}} {\bf{\Lambda }} {{\bf{V}}^{-T}} \nonumber \\
\label{err6}
 & \;\;\;\;\; - {{\bf{V}}^{-1}} {\bf{\Lambda }} {{\bf{V}}^{-T}} {{\bf{U}}^T}{\left( {{\bf{\Lambda }} + {{\bf{\Sigma }}_{\bf{\nu}}}} \right)^{ - 1}}{\bf{U}} {{\bf{V}}^{-1}} {\bf{\Lambda }} {{\bf{V}}^{-T}} \\
 &= {{\bf{\Sigma }}_{\bf{x|yz}}} + {{\bf{V}}^{-1}} {\bf{\Lambda }} {{\bf{V}}^{-T}} \nonumber \\
\label{err7}
 & \;\; - \!{{\bf{V}}^{-1}} \! {\bf{\Lambda }} {\left( \!\! {{\bf{\Lambda }} \!+\! {\rm{diag}} \! \left\{\!\! { \frac{{\lambda _i}\min \left({\lambda _i},{\lambda' _i}\right)}{{\lambda _i} \!-\! \min \left({\lambda _i},{\lambda' _i} \right)} } , i \! = \! 1,\dotsc, \! n_x \!\! \right\} } \!\! \right)^{ - 1}} \! {\bf{\Lambda }} \! {{\bf{V}}^{-T}} \\
&= {{\bf{\Sigma }}_{\bf{x|yz}}} + {{\bf{V}}^{-1}} {\rm{diag}}\left\{ {{\min \left({\lambda _i},{\lambda' _i} \right)} } , i=1,\dotsc,n_x \right\} {{\bf{V}}^{-T}} \nonumber \\
\label{err9}
&= {{\bf{\Sigma }}_{\bf{x|yz}}} + {\min \left( {{\bf{\Sigma }}_{\bf{x|z}}} - {{\bf{\Sigma }}_{\bf{x|yz}}} , {\bf{D}} - {{\bf{\Sigma }}_{\bf{x|yz}}} \right)}
\end{align}

\noindent
where (\ref{err1}) and (\ref{err2}) follow from (\ref{xuz}), (\ref{err3}) is the result of substituting (\ref{weird1}) in (\ref{err2}), (\ref{err4}) follows from (\ref{weird2}) and (\ref{weird6}), (\ref{err6}) follows from (\ref{joint1}), and (\ref{err7}) follows from (\ref{ub6}) and (\ref{cov_noise}).

Finally, rewriting (\ref{err9}) as:

\begin{equation}
\label{D}
{{\bf{\Sigma }}_{{\bf{x}}|{\bf{u^*}}{{\bf{z}}}}} - {{\bf{\Sigma }}_{{\bf{x}}|{\bf{y}}{{\bf{z}}}}} = {\min \left( {{\bf{\Sigma }}_{\bf{x|z}}} - {{\bf{\Sigma }}_{\bf{x|yz}}} , {\bf{D}} - {{\bf{\Sigma }}_{\bf{x|yz}}} \right)},
\end{equation}

\noindent
and using Property 3 in Lemma \ref{lem_properties}, it is clear that ${{\bf{\Sigma }}_{\bf{x|u^*z}}} \preceq {\bf{D}}$, as desired. This completes the proof.

%********************************************************************************************************************************************************
\section{Derivation of the KKT Conditions}
\label{appendix_kkt}

We start with defining ${\bf{A}}$ as in (\ref{KKT_result2}). Writing ${\bf{D}}_i$ in terms of ${\bf{Z}}_i$ using (\ref{KKT1}), and substituting in the constraint in (\ref{MIN2}) yields (\ref{KKT_result3}). We derive (\ref{KKT_result1}) to complete the proof. Applying the matrix inversion lemma to ${\left[  {{\bf{\Sigma}}_{{\bf{n}}_i} } + {\left( {\bf{D}}_i^{-1} - {\bf{\Sigma}}_{{\bf{y}}_i}^{-1} \right)}^{-1}  \right]}^{-1}$ in (\ref{MIN2}), and using the definitions (\ref{KKT1}) and (\ref{KKT_result2}), we rewrite the optimization problem as follows:

\begin{align}
&\min_{{\bf{Z}}_1, \cdots, {\bf{Z}}_N} {\text{tr}  \left\{  {\bf{A}} ^{-1}  \right\}} \nonumber \\
\label{MIN3}
&   \;\;\;\; s.t. \;\;\;\;  \prod_{i=1}^N \left| {\bf{\Sigma}}_{{\bf{y}}_i}^{-1} - {\bf{\Sigma}}_{{\bf{n}}_i}^{-1} + {\bf{\Sigma}}_{{\bf{n}}_i}^{-1} {\bf{W}}_i^T {\bf{Z}}_i^{-1} {\bf{W}}_i {\bf{\Sigma}}_{{\bf{n}}_i}^{-1} \right| ^{- \alpha_i} = \beta.
\end{align}

\noindent
For a given $i \in \{ 1,\cdots,N\}$, we differentiate the Lagrangian of (\ref{MIN3}) with respect to ${\bf{Z}}_i$. To do so, we first apply a logarithm function to the constraint in (\ref{MIN3}), and then write the Lagrangian form of the problem as:

\begin{equation}
\label{Lagrange}
{\mathcal{L}} ( \!\lambda , {\bf{Z}}_1,\cdots,{\bf{Z}}_N \!) \!=\! {\text{tr} \! \left\{\! \left( {\bf{A}}_i  - {\bf{Z}}_i \right)^{-1}  \!\right\}} + \lambda \!\!\left(\! \sum_{j=1}^N {f_j({\bf{Z}}_j)} \!-\! \log{\beta} \!\!\right)\!\!,
\end{equation}
 
\noindent
where ${\bf{A}}_i$ and $f_j({\bf{Z}}_j)$ are defined as:

\begin{align*}
& {\bf{A}}_i = {\bf{A}} +  {\bf{Z}}_i, \\
& f_j({\bf{Z}}_j) = -\alpha_j \log{ \left| {\bf{\Sigma}}_{{\bf{y}}_j}^{-1} - {\bf{\Sigma}}_{{\bf{n}}_j}^{-1} + {\bf{\Sigma}}_{{\bf{n}}_j}^{-1} {\bf{W}}_j^T {\bf{Z}}_j^{-1} {\bf{W}}_j {\bf{\Sigma}}_{{\bf{n}}_j}^{-1} \right| }.
\end{align*}

\noindent
Note that from (\ref{KKT_result2}) it follows that ${\bf{A}}_i$ does not depend on ${\bf{Z}}_i$. Using (\ref{KKT2}), we rewrite $f_j({\bf{Z}}_j)$ as:

\begin{align}
f_j({\bf{Z}}_j) = &-\alpha_j \log{ \left|  {\bf{\Sigma}}_{{\bf{n}}_j}^{-1} {\bf{W}}_j^T \left( {\bf{Z}}_j^{-1} - {\bf{C}}_j^{-1} \right) {\bf{W}}_j {\bf{\Sigma}}_{{\bf{n}}_j}^{-1}  \right|} \nonumber \\
 = &-\alpha_j \log{ \left|  {\bf{\Sigma}}_{{\bf{n}}_j}^{-1} {\bf{W}}_j^T {\bf{W}}_j {\bf{\Sigma}}_{{\bf{n}}_j}^{-1}  \right|} -\alpha_j \log{ \left| {\bf{C}}_j^{-1} \right|} \nonumber \\
\label{fj2}
 & -\alpha_j \log{ \left| {\bf{Z}}_j^{-1} \right|} -\alpha_j \log{ \left| {\bf{C}}_j - {\bf{Z}}_j \right|}.
\end{align}

\noindent
Substituting (\ref{fj2}) in (\ref{Lagrange}), differentiating with respect to ${\bf{Z}}_i$ while taking into account that ${\bf{Z}}_i$ is symmetric \cite{cook}, and setting the derivative equal to $0$ yields:

\begin{align}
& 2 \!\left( \!{\bf{A}}_i  \!-\! {\bf{Z}}_i \right)^{-2} \!-\! \left( {\bf{A}}_i  \!-\! {\bf{Z}}_i \right)^{-2} \!\circ\! {\bf{I}}_n + 2\lambda \alpha_i \!\left[ {\bf{Z}}_i^{-1} \!+\! \left( {\bf{C}}_i \!-\! {\bf{Z}}_i \right)^{-1} \right] \nonumber \\
\label{deriv_hadamard}
& - \lambda \alpha_i \left[ {\bf{Z}}_i^{-1} + \left( {\bf{C}}_i - {\bf{Z}}_i \right)^{-1} \right] \circ {\bf{I}}_n = {\bf{0}},
\end{align}

\noindent
where $\circ$ denotes the Hadamard product. From (\ref{deriv_hadamard}) it follows that:

\begin{equation}
\label{deriv}
\left( {\bf{A}}_i  - {\bf{Z}}_i \right)^{-2} + \lambda \alpha_i \left[ {\bf{Z}}_i^{-1} + \left( {\bf{C}}_i - {\bf{Z}}_i \right)^{-1} \right] = {\bf{0}}.
\end{equation}

\noindent
Replacing ${\bf{A}}_i  - {\bf{Z}}_i$ by ${\bf{A}}$ in (\ref{deriv}) and simplifying the result yields (\ref{KKT_result1}).

%********************************************************************************************************************************************************
\section{Proof of Proposition \ref{high_rate_proposition}}
\label{appendix_uniqueness_highrate}

Let us define: 

\begin{equation*}
f_i(\lambda) = \left( \frac{\sqrt{1+4 \lambda s_i} - 1}{\sqrt{4 \lambda}} \right)^2; i=1,\cdots,N.
\end{equation*}

Using basic calculus one can show that $f_i(\lambda)$ is a monotonically increasing and strictly concave function of $\lambda$, which varies from $0$ to $s_i$ when $\lambda$ goes from $0$ to $\infty$. It then follows that the function $f(\lambda) = \prod_{i=1}^N {f_i(\lambda)} - \gamma$ is monotonically increasing with the range $[-\gamma , \left| {\bf{S}} \right| - \gamma [$. One could then see that $f(\lambda)$ has one root, if and only if:

\begin{equation}
\label{monotone_root1}
\left| {\bf{S}}  \right| - \gamma > 0,
\end{equation}

\noindent
otherwise it has no root. Substituting (\ref{gamma}) in (\ref{monotone_root1}) and simplifying the result yields $R > R_{min}$.

%********************************************************************************************************************************************************
\section{Proof of Proposition \ref{prop}}
\label{appendix_scalar}

Applying the assumptions $\alpha_1 = \alpha_2 = 0.5$, $w_1 = w_2$, and $n = 1$, the optimization problem is simplified to:

\begin{align}
\max_{D_1, D_2} & \frac{1}{\Sigma_{n_1}} - \frac{1}{{\Sigma_{n_1}}^2} \left( \frac{1}{ \frac{1}{\Sigma_{n_1}} + \frac{1}{D_1} - \frac{1}{\Sigma_{y_1}} } \right) \nonumber \\
\label{scalar1}
& + \frac{1}{\Sigma_{n_2}} \!-\! \frac{1}{{\Sigma_{n_2}}^2} \!\left(\! \frac{1}{ \frac{1}{\Sigma_{n_2}} + \frac{1}{D_2} - \frac{1}{\Sigma_{y_2}} } \!\right)  \;\; s.t. \;\;  D_1 D_2 = \beta^2. 
\end{align}

\noindent
%Removing the fixed terms $\frac{1}{\Sigma_{n_1}}$ and $\frac{1}{\Sigma_{n_2}}$ and using 
Applying the following change of variables:

\begin{equation}
\label{change_var}
D'_i = \Sigma_i + \frac{{\Sigma_{n_i}}^2}{D_i}, \;\;\;\; i=1,2,
\end{equation}

\noindent
and defining $\beta'$ as:

\begin{equation}
\label{beta}
\beta' = (\Sigma_{n_1} - \Sigma_1)(\Sigma_{n_2} - \Sigma_2)e^{4R}, 
\end{equation}

\noindent
one could rewrite (\ref{scalar1}) as:

\begin{equation}
\label{scalar2}
\min_{D'_1, D'_2} \left( \frac{1}{D'_1} + \frac{1}{D'_2} \right)  \;\; \;\; s.t. \;\; \;\;  (D'_1 - \Sigma_1)(D'_2 - \Sigma_2) = \beta'.
\end{equation}

\noindent
Using the constraint in (\ref{scalar2}), we then write $D'_2$ in terms of $D'_1$ and substitute it in the cost function to obtain an unconstrained function of $D'_1$. We call this function $f_1(D'_1)$. Differentiating $f_1(D'_1)$ with respect to $D'_1$ yields:

\begin{align}
&\frac{df_1(D'_1)}{dD'_1} = \nonumber \\
\label{diff}
&\frac{\left[ \sqrt{\beta'} D'_1 \!+\! \beta' \!+\! \Sigma_2 (D'_1  \!-\! \Sigma_1) \right] \left[ (\sqrt{\beta'} \!-\! \Sigma_2)D'_1 \!-\! (\beta' \!-\! \Sigma_1 \Sigma_2) \right]}{\left( D'_1 \right)^2 \left( \Sigma_2 D'_1 \!-\! \Sigma_1 \Sigma_2 \!+\! \beta' \right)^2}, 
\end{align}

\noindent
which is zero at $D'^*_1 = \Sigma_1 + \frac{{\Sigma_{n_1}}^2}{D^*_1}$ (equivalent to $D_1 = D^*_1$ using (\ref{change_var})). The sign of the derivative around this point determines whether there is an extremum or not. The denominator and the first term in the numerator in (\ref{diff}) are positive. We thus study only the second term in the numerator. We replace $D'_1$ by $\Sigma_1 + \frac{{\Sigma_{n_1}}^2}{D^*_1} + \epsilon$. The result is $(\sqrt{\beta'} - \Sigma_2) \epsilon$, which implies that if $\sqrt{\beta'} > \Sigma_2$, the derivative is positive for $\epsilon > 0$ and negative for $\epsilon < 0$. This means that if $\sqrt{\beta'} > \Sigma_2$, the minimum is at $D'_1 = \Sigma_1 + \frac{{\Sigma_{n_1}}^2}{D^*_1}$, or equivalently $D_1 = D^*_1$ is the global maximizer of the SNR. Similarly, one could write $D'_1$ in terms of $D'_2$ and substitute in the cost function to obtain $f_2(D'_2)$. Considering the sign of the derivative around $D_2 = D^*_2$ leads to the conclusion that if $\sqrt{\beta'} > \Sigma_1$, the SNR is maximized at $D_2 = D^*_2$. Combining the two conditions, the following should hold in order to have the SNR maximized at $( D^*_1,D^*_2)$:
\vspace{-1em}

\begin{equation}
\label{max1}
\sqrt{\beta'} > \max (\Sigma_1,\Sigma_2).
\end{equation}

\noindent
Substituting (\ref{beta}) in (\ref{max1}) yields:

\begin{equation}
\label{max2}
R > \frac{1}{4} \log \frac{\left[\max(\Sigma_1,\Sigma_2)\right]^2}{(\Sigma_{n_1} - \Sigma_1)(\Sigma_{n_2} - \Sigma_2)} = R_{max}.
\end{equation}

Similarly one can show that for $R < R_{min}$, $( D^*_1,D^*_2)$ minimizes the output SNR. This completes the proof.

%********************************************************************************************************************************************************
%********************************************************************************************************************************************************
%********************************************************************************************************************************************************
%********************************************************************************************************************************************************

\begin{IEEEbiography}[{\includegraphics[width=1in,height=1.25in,clip,keepaspectratio]{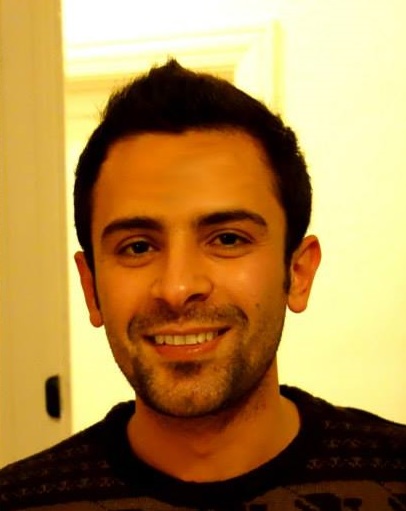}}]{Adel Zahedi}
 received the M.Sc. degree from Iran University of Science and Technology, Iran in 2011, and the Ph.D. degree from Aalborg University, Denmark in 2016. He is currently a postdoctoral researcher at Aalborg University.
\end{IEEEbiography}

\begin{IEEEbiography}[{\includegraphics[width=1in,height=1.25in,clip,keepaspectratio]{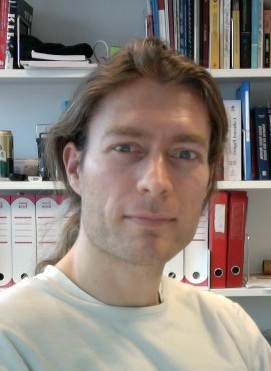}}]{Jan {\O}stergaard}
(S'98--M'99--SM'11) received the M.Sc.E.E.\ degree from Aalborg University, Aalborg, Denmark, in 1999 and the Ph.D.\ degree (\textit{cum laude}) from Delft University of Technology, Delft, The Netherlands, in 2007.

From 1999 to 2002, he worked as an R\&D Engineer at ETI A/S, Aalborg, and from 2002 to 2003, he was an R\&D Engineer with ETI Inc., VA, USA.
Between September 2007 and June 2008, he was a Postdoctoral Researcher at The University of Newcastle, NSW, Australia.
From June 2008 to March 2011, he was a Postdoctoral Researcher/Assistant Professor at Aalborg University, and, since 2011, has been an Associate Professor with the same University.
He has been a Visiting Researcher at Tel Aviv University, Tel Aviv, Israel, and at the Universidad T{\'e}cnica Federico Santa Mar{\'i}a, Valpara{\'i}so, Chile.

Dr.\ {\O}stergaard received a Danish Independent Research Council's Young Researcher's Award, a Best Ph.D.\ Thesis award by the European Association for Signal Processing (EURASIP), and fellowships from the Danish Independent Research Council and the Villum Foundation's Young Investigator Programme.
He is an Associate Editor of the EURASIP Journal on \textit{Advances in Signal Processing}
\end{IEEEbiography}

\begin{IEEEbiography}[{\includegraphics[width=1in,height=1.25in,clip,keepaspectratio]{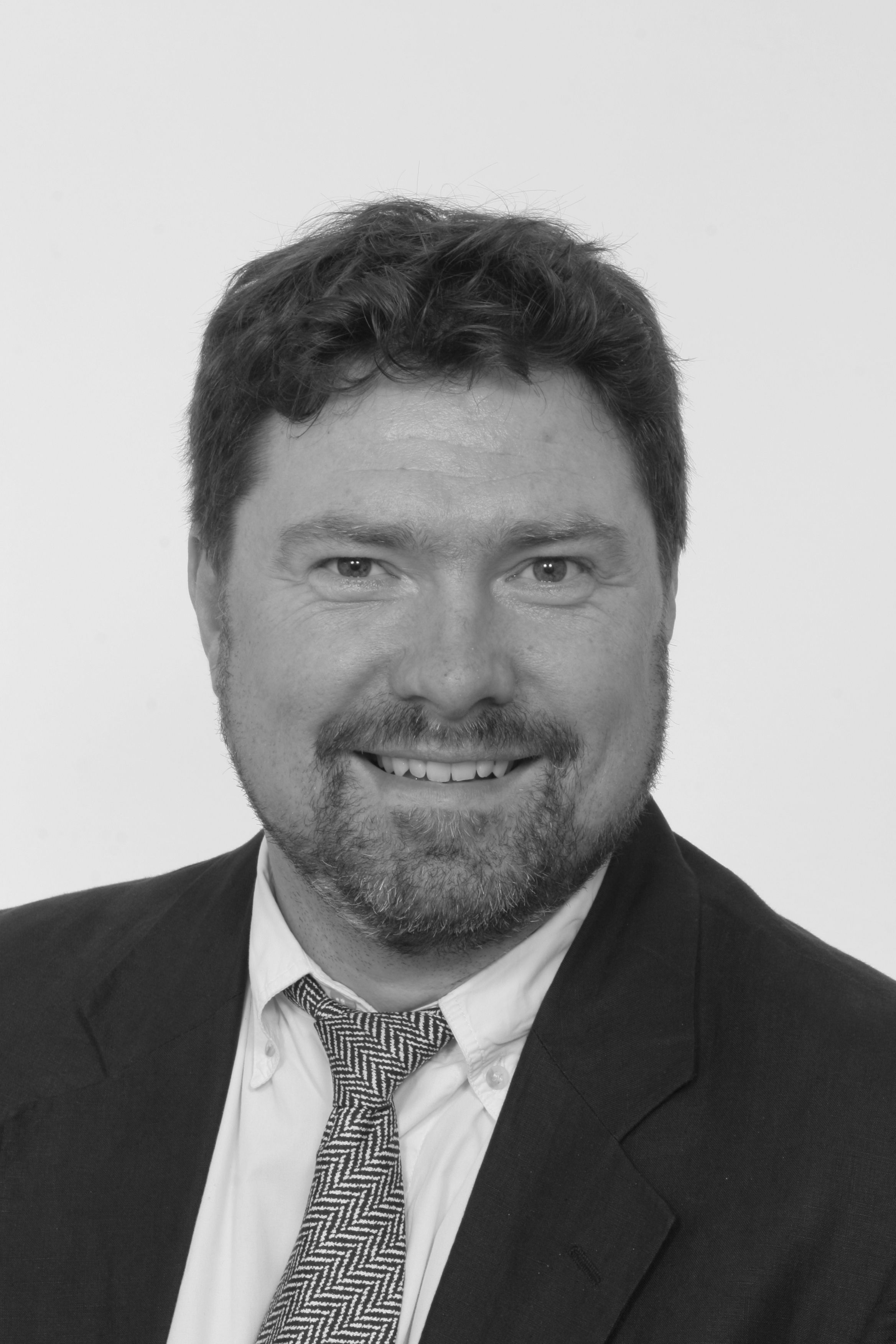}}]{S\o ren Holdt Jensen}
 (S'87-M'88-SM'00) received the M.Sc.\ degree in 
electrical engineering from Aalborg University, Aalborg, Denmark, in 
1988, and the Ph.D.\ degree in signal processing from the Technical 
University of Denmark, Lyngby, Denmark, in 1995. Before joining the 
Department of Electronic Systems of Aalborg University, he was with the 
Telecommunications Laboratory of Telecom Denmark, Ltd, Copenhagen, 
Denmark; the Electronics Institute of the Technical University of 
Denmark; the Scientific Computing Group of Danish Computing Center for 
Research and Education (UNI{\textbullet}C), Lyngby; the Electrical 
Engineering Department of Katholieke Universiteit Leuven, Leuven, 
Belgium; and the Center for PersonKommunikation (CPK) of Aalborg 
University. He is Full Professor and heading a research team working in 
the area of numerical algorithms, optimization, and signal processing 
for speech and audio processing, image and video processing, multimedia 
technologies, and digital communications. Prof.\ Jensen was an Associate 
Editor for the IEEE Transactions on Signal Processing, Elsevier Signal 
Processing and EURASIP Journal on Advances in Signal Processing, and is 
currently Associate Editor for the IEEE/ACM Transactions on Audio, 
Speech and Language Processing. He is a recipient of an European 
Community Marie Curie Fellowship, former Chairman of the IEEE Denmark 
Section and the IEEE Denmark Section's Signal Processing Chapter. He is 
member of the Danish Academy of Technical Sciences and was in January 
2011 appointed as member of the Danish Council for Independent 
Research---Technology and Production Sciences by the Danish Minister for 
Science, Technology and Innovation.
\end{IEEEbiography}

\begin{IEEEbiography}[{\includegraphics[width=1in,height=1.25in,clip,keepaspectratio]{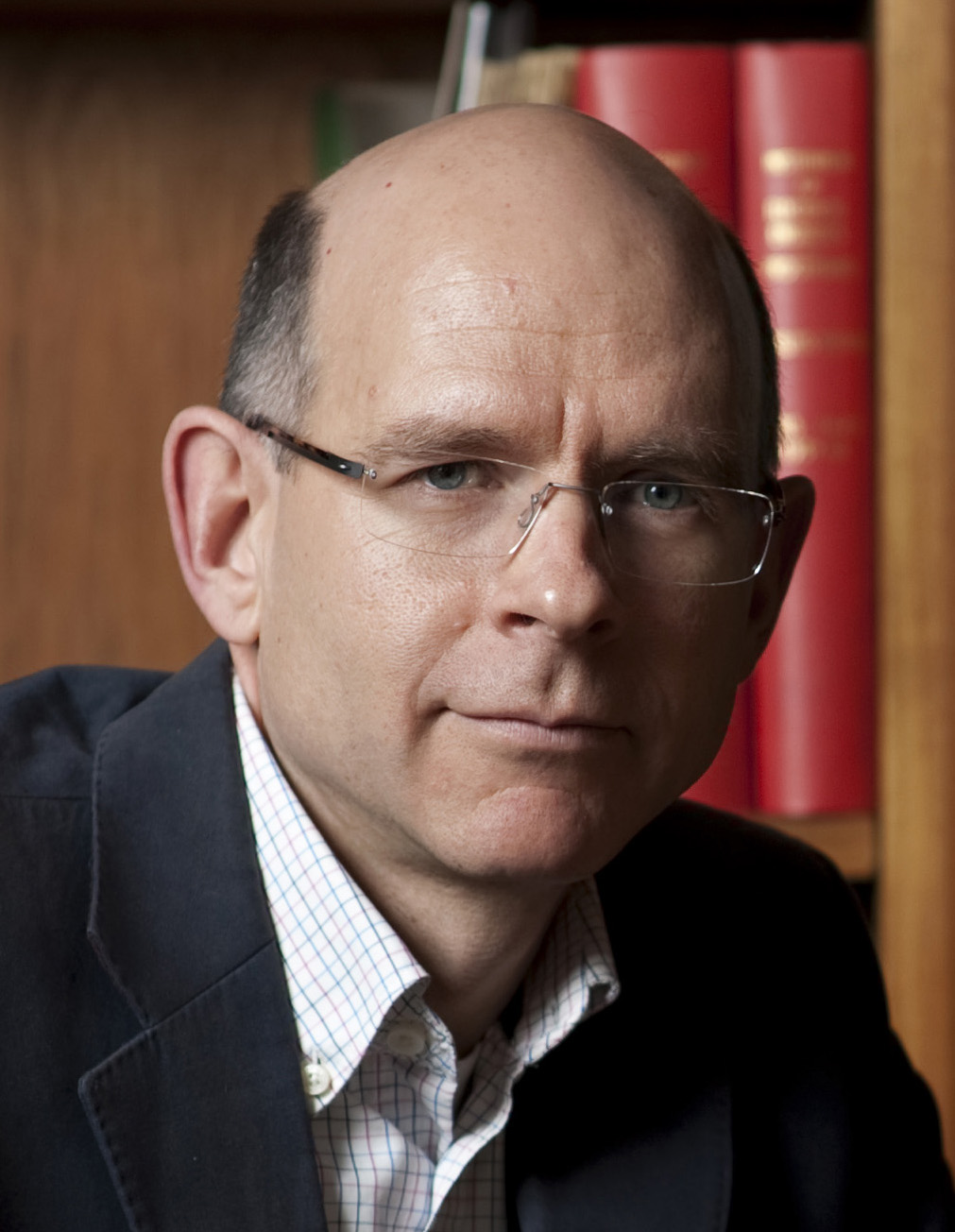}}]{Patrick Naylor}
 (M'89, SM'07) received his BEng degree in Electronic and Electrical Engineering from the University of Sheffield, U.K., in 1986 and the PhD. degree from Imperial College, London, U.K., in 1990. Since 1990 he has been a member of academic staff in the Department of Electrical and Electronic Engineering at Imperial College London. His research interests are in the areas of speech, audio and acoustic signal processing. He has worked in particular on adaptive signal processing for dereverberation, blind multichannel system identification and equalization, acoustic echo control, speech quality estimation and classification, single and multi-channel speech enhancement and speech production modelling with particular focus on the analysis of the voice source signal. In addition to his academic research, he enjoys several fruitful links with industry in the UK, USA and in mainland Europe. He is the Chair of the IEEE Signal Processing Society Technical Committee on Audio and Acoustic Signal Processing, a director of the European Association for Signal Processing (EURASIP) and formerly an associate editor of IEEE Signal Processing Letters and IEEE Transactions on Audio Speech and Language Processing.
\end{IEEEbiography}

\begin{IEEEbiography}[{\includegraphics[width=1in,height=1.25in,clip,keepaspectratio]{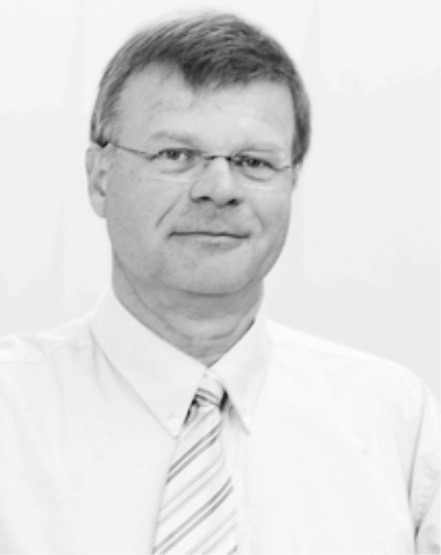}}]{S\o ren Bech}
 received a M.Sc. and a Ph.D. from the Department of Acoustic Technology (AT) of the Technical University of Denmark. From 1982 to 1992 he was a research Fellow at AT studying perception and evaluation of reproduced sound in small rooms. In 1992 he joined Bang \& Olufsen where he is currently Head of Research. In 2011 he was appointed Professor in Audio Perception at Aalborg University. His research interest includes human perception of reproduced sound in small and medium sized rooms. experimental procedures and statistical analysis of data from sensory analysis of audio and video quality. General perception of sound in small rooms is also a major research interest. 
\end{IEEEbiography}


%\end{IEEEbiography}
%% insert where needed to balance the two columns on the last page with
%%\newpage


%\vfill


\end{document}